\documentclass[a4paper,11pt]{article}
\pdfoutput=1

\usepackage{jheppub}

\makeatletter
\gdef\@fpheader{}
\makeatother
\usepackage[T1]{fontenc}
\usepackage{lmodern}
\usepackage{booktabs}
\usepackage{array}
\usepackage{bm}
\usepackage{microtype}
\usepackage{tikz}
\usepackage{colortbl}
\usepackage{float}
\definecolor{tabhead}{RGB}{214,220,250}
\definecolor{tabrow}{RGB}{238,241,253}
\usetikzlibrary{arrows.meta,calc,decorations.pathmorphing,decorations.markings}

\tikzset{
  fermion/.style={draw, postaction={decorate},
    decoration={markings, mark=at position 0.55 with {\arrow{Stealth[length=2mm]}}}},
  heavyfermion/.style={draw, double, double distance=0.7pt,
    postaction={decorate},
    decoration={markings, mark=at position 0.55 with {\arrow{Stealth[length=2mm]}}}},
  scalarline/.style={draw, dashed, dash pattern=on 2.4pt off 2.2pt},
  boson/.style={draw, decorate,
    decoration={coil, aspect=0.6, segment length=2.1mm, amplitude=0.85mm}},
  vertexdot/.style={circle, fill, inner sep=1.05pt},
}

\graphicspath{{figures/}}

\definecolor{changedcolor}{RGB}{0,0,0}
\newcommand{\changed}[1]{{\color{changedcolor}#1}}
\definecolor{revred}{RGB}{190,0,0}

\usepackage{placeins}

\hypersetup{
  pdftitle={Testing Minimal U(2)^3 Right-Handed Flavor Breaking in a
            Singlet-Scalar Vector-Like Quark Portal},
  pdfauthor={Qazi Maaz Us Salam},
  pdfsubject={Flavor physics, vector-like quarks, neutral-meson mixing},
  pdfkeywords={Beyond Standard Model; Flavor Physics; CP violation;
               Vector-like quarks},
}

\newcommand{\GF}{G_F}

\newcommand{\epsR}{\epsilon_R}
\newcommand{\thetaR}{\theta_R}

\newcommand{\Cseven}{C_{7\gamma}}
\newcommand{\Ceight}{C_{8G}}
\newcommand{\CRR}{C_{RR}}
\newcommand{\BR}{\mathcal B}
\newcommand{\cL}{\mathcal L}
\newcommand{\cH}{\mathcal H}
\newcommand{\cO}{\mathcal O}
\newcommand{\NP}{\mathrm{NP}}
\newcommand{\SM}{\mathrm{SM}}
\newcommand{\TeV}{\,\mathrm{TeV}}
\providecommand{\slashed}[1]{\ensuremath{\not\mathrel{\mkern-1mu}#1}}
\title{\boldmath Testing Minimal $U(2)^3$ Right-Handed Flavor Breaking in a
Singlet-Scalar Vector-Like Quark Portal}

\author{Qazi Maaz Us Salam}
\affiliation{Department of Physics, Lahore University of Management
Sciences (LUMS), Opposite Sector U, D.H.A, Lahore 54792, Pakistan}
\emailAdd{qazimaaz92@gmail.com}

\abstract{%
\changed{We study a portal of a $Z_2$-odd singlet scalar and a vector like down quark, in which a right handed flavor breaking spurion extends the portal coupling from $b_R$ to $d_R$ and $s_R$. This single coupling vector links the $b\to s$, $b\to d$, and $s\to d$ transitions. We compute the one loop dipole and $\Delta F=2$ Wilson coefficients, and show that a real scalar generates no $\Delta F=2$ amplitude at one loop. Using a seven observable profile likelihood of $B$ meson mixing, $\epsilon_K$, and radiative decays, with the CKM, lattice, and short distance inputs treated as nuisance parameters, we compare a generic spurion, minimal $U(2)^3$ with a free size of the light quark couplings, and a CKM-aligned ansatz. We have found that the present data cannot yet test the flavor structure: the Standard Model is compatible with the data, with a pseudo experiment $p$-value of $0.20$, and no direction of the light quark couplings can be excluded at 95\% CL. For $M_D\simeq m_{\mathcal S}=1\TeV$ and $|\lambda_b|=1$, the relative light quark coupling $\epsilon_R$ is bounded at 95\% CL by $0.227$ for a generic spurion and by $0.211$ for minimal $U(2)^3$. More precise measurements of the $B_s$ mixing phase at the LHC and of $\sin2\beta$ at LHCb and Belle~II, together with improved lattice and CKM inputs, can make the minimality of the flavor structure testable if a deviation from the Standard Model is observed.}}

\begin{document}
\maketitle
\flushbottom

\section{Introduction}
\label{sec:introduction}

Flavor-changing neutral current (FCNC) processes do not occur at tree level in the Standard Model (SM). They arise only at the loop level, where they are further suppressed by the Glashow-Iliopoulos-Maiani (GIM) mechanism \cite{Glashow:1970gm}. For this reason, FCNC processes are sensitive to new heavy particles even when these particles cannot be produced directly, and the resulting bounds on generic flavor violation reach scales far above those probed at colliders \cite{Isidori:2010kg,Buras:2013ooa,Silvestrini:2018dos}. Among these processes, radiative decays, which change flavor by one unit ($\Delta F=1$), and neutral-meson mixing, which changes it by two units ($\Delta F=2$), are particularly useful, since they depend on the same new couplings in different ways; their current experimental and theoretical status is reviewed in refs.~\cite{Albrecht:2024oyn,ParticleDataGroup:2026mpi,Blake:2016olu,Artuso:2015swg}. In previous work, FCNC effects were studied in several channels \cite{Salam:2026vys,Salam:2024nfv,Aarfi:2025qcp,Aarfi:2026hgi,Aarfi:2026cea}.

The sensitivity of FCNC processes to new physics (NP), however, also poses a challenge. If new particles exist at the TeV scale, as many extensions of the SM suggest \cite{Nilles:1983ge,Arkani-Hamed:1998jmv,Randall:1999ee,Arkani-Hamed:2001nha,Agashe:2004rs,Bellazzini:2014yua,Craig:2022eqo,Alves:2023ufm,Banerjee:2024zvg}, and couple to quarks with a generic flavor structure, they would produce FCNC effects far larger than observed. This is the so-called NP flavor problem, which can be avoided either by making the new particles very heavy or by relating their flavor couplings to those of the SM \cite{Isidori:2010kg}. Moreover, in a generic extension of the SM the new couplings entering the $b\to s$, $b\to d$, and $s\to d$ transitions are independent of each other. Each process then constrains a different coupling, and the separate bounds reveal little about the underlying flavor structure. Minimal Flavor Violation (MFV) \cite{Chivukula:1987py,Buras:2000dm,DAmbrosio:2002vsn} is the most restrictive way of relating the new couplings to those of the SM. In MFV, the SM Yukawa matrices are assumed to be the only sources of flavor breaking, so that every new flavor-violating amplitude inherits the hierarchy of the CKM matrix. However, MFV is based on a $U(3)^3$ flavor symmetry \cite{DAmbrosio:2002vsn,Barbieri:2011ci} that treats all three generations alike, and this symmetry is strongly broken by the large top Yukawa coupling \cite{Kagan:2009bn}. It is therefore natural to consider instead a smaller $U(2)^3$ symmetry acting only on the first two generations \cite{Barbieri:2011ci,Barbieri:2012uh,Barbieri:2014tja}, which is the framework of the present work.

The $U(2)^3$ flavor symmetry
\begin{equation}
G_F=U(2)_Q\times U(2)_u\times U(2)_d
\label{eq:flavor-group}
\end{equation}
acts on the first two quark generations, while the third generation is a singlet \cite{Barbieri:2011ci,Barbieri:2012uh}. As a result, new flavor-violating effects among the first two generations, such as those in kaon mixing, remain as strongly suppressed as in MFV. The third generation, on the other hand, is not restricted by the symmetry, so that NP can couple strongly to the $b$ and $t$ quarks. This pattern is consistent with the data, since experiments allow much larger new effects in transitions involving the $b$ quark, such as $B_s$ mixing, than in transitions involving only the $d$ and $s$ quarks, such as kaon mixing \cite{Isidori:2010kg,Silvestrini:2018dos}. It is also expected in many extensions of the SM, in which new particles couple most strongly to the third generation because it contains the heavy top quark \cite{Barbieri:2011ci,Bellazzini:2014yua}. In the minimal version of $U(2)^3$, the observed quark masses and CKM mixing are reproduced by three symmetry-breaking spurions, $\Delta Y_u$, $\Delta Y_d$, and $V$ \cite{Barbieri:2012uh,Barbieri:2012bh}. The phenomenology of $U(2)$ flavor symmetries has been studied extensively \cite{Barbieri:2014tja,Fuentes-Martin:2019mun,Crivellin:2019isj,Faroughy:2020ina,Greljo:2022cah}. A more general version of $U(2)^3$ also allows an additional right-handed spurion $V_d$, which generates right-handed flavor violation that is not fixed by the left-handed spurion $V$ \cite{Barbieri:2012bh}. This distinction is essential for the present work, since the portal considered here couples directly to the right-handed quarks $d_R$, $s_R$, and $b_R$.

The portal considered here consists of a neutral singlet scalar $\mathcal S$ with mass $m_{\mathcal S}$ and a vector-like down quark $D\sim(3,1,-1/3)$ with mass $M_D$, the same field content as in ref.~\cite{Salam:2026vys}. The mixing patterns and production modes of vector-like quarks are given in ref.~\cite{Aguilar-Saavedra:2013qpa}, and recent reviews are given in refs.~\cite{Alves:2023ufm,Banerjee:2024zvg}. The same field content also appears in $t$-channel dark-matter models, in which the singlet scalar is the dark-matter candidate \cite{Giacchino:2015hvk,Baek:2017ykw,Biondini:2019int,Arina:2020tuw,Becker:2022iso,Arina:2023msd,Arina:2025zpi}. These studies focus on the relic abundance, direct detection, and collider searches, mostly with couplings to a single quark flavor, whereas the present work concerns the down-type flavor-changing transitions; the dark-matter implications of our benchmark are briefly discussed in section~\ref{sec:nonflavor}.

Flavor-changing couplings of a $Z_2$-odd sector to quarks have also been studied in the framework of Dark Minimal Flavor Violation (DMFV) \cite{Agrawal:2014aoa}. In DMFV, a flavor triplet of dark fermions couples to the quarks through a colored scalar mediator, and the $3\times3$ coupling matrix is the only new source of flavor violation. Subsequent analyses examined top-flavored and left-handed versions of this framework, combining meson mixing with relic-density, direct-detection, and collider constraints \cite{Blanke:2017tnb,Blanke:2017fum}, and the broader phenomenology of flavored dark matter and of dark matter with $t$-channel mediators is reviewed in refs.~\cite{Kile:2013ola,Garny:2015wea}. Our portal differs from these models in two ways. First, it contains a single scalar with only three couplings, one to each of the $d$, $s$, and $b$ quarks, instead of a $3\times3$ coupling matrix. As a result, the $b\to s$, $b\to d$, and $s\to d$ transitions are not independent, and the measurements in the $B_s$, $B_d$, and kaon systems test one common flavor structure. Second, the particle in the loop that connects the scalar to the quarks is a colored fermion, the vector-like quark $D$, rather than a colored scalar. The one-loop formulas for such loops \cite{Bobeth:1999ww,Arnan:2019uhr} and the relevant $\Delta F=2$ operators \cite{Gabbiani:1996hi} are known in general, and we use them. Beyond constraining the size of these couplings, we test whether the ratio of the couplings to the $d$ and $s$ quarks agrees with the prediction of minimal $U(2)^3$.

This field content was first studied in ref.~\cite{Salam:2026vys}, in a two-flavor setup in which the portal couples only to the $s$ and $b$ quarks. That work established three main components. First, since the scalar is electrically neutral and a color singlet, the photon and the gluon can be emitted only from the internal $D$ line, which fixes the ratio of the NP dipole Wilson coefficients (WCs) to $\Cseven^{\NP}/\Ceight^{\NP}=Q_D=-1/3$. Second, since the portal couples to quarks of only one chirality, the dipole amplitude is not enhanced by the large ratio $M_D/m_b$, where $m_b$ is the $b$-quark mass, so that $B\to X_s\gamma$ gives only a weak constraint. Third, in $B_s$ mixing a complex scalar produces a right-handed box contribution, whereas for a single real scalar the direct and crossed boxes cancel exactly, so that $B_s$ mixing gives the stronger constraint only for the complex scalar.

The purpose of the present work is to extend that benchmark to all three generations and to use it as a test of the flavor structure. As discussed above, the single coupling vector fixes the $s\to d$ transition once the $b\to d$ and $b\to s$ transitions are known, independently of any flavor symmetry. What a flavor symmetry can add is a prediction for the ratio of the $b\to d$ and $b\to s$ couplings, and we compare three possibilities. A generic right-handed spurion leaves this ratio free \cite{Barbieri:2012bh}, so that no prediction arises. In minimal $U(2)^3$, the right-handed spurion must be built from the spurions that generate the SM Yukawa couplings, and the simplest such combination, $\Delta Y_d^\dagger V$, carries the down-quark Yukawa couplings. The $b\to d$ coupling is therefore suppressed relative to the $b\to s$ coupling by an additional factor $m_d/m_s$. For contrast, we also consider a CKM-aligned ansatz, in which the portal couplings follow the CKM matrix without this additional factor. This ansatz is what one would write by analogy with the left-handed sector, but it is not a consequence of minimal $U(2)^3$. The two predictive structures can be compared through the ratio $h_d/h_s$ of the NP contributions to $B_d$ and $B_s$ mixing: minimal $U(2)^3$ predicts $h_d/h_s=(m_d/m_s)^2\simeq2.5\times10^{-3}$, whereas the CKM-aligned ansatz predicts $h_d/h_s=1$. The two structures therefore differ by more than two orders of magnitude and lead to different signatures. In minimal $U(2)^3$, an NP effect should appear almost only in $B_s$ mixing, whereas in the CKM-aligned case it should appear equally in $B_d$ and $B_s$ mixing. The central aim of this paper is to confront these three structures with the current meson-mixing and radiative-decay measurements, and in particular to determine whether the data require a flavor structure beyond minimal $U(2)^3$.

Besides the flavor structure, the pattern of constraints also depends on whether the $Z_2$-odd scalar is complex or real. For a complex scalar, the $\Delta F=2$ amplitudes depend on the squares of the coupling products, so that $K$, $B_d$, and $B_s$ mixing constrain one common flavor structure and are sensitive to twice the phases of the couplings. For a single real scalar, the leading box contributions to $K$, $B_d$, and $B_s$ mixing cancel exactly at one loop, and only the radiative decays $b\to s\gamma$ and $b\to d\gamma$ remain. As we show below, these decays are too weak to constrain the couplings at the one-TeV benchmark, so that the two cases lead to qualitatively different results.

The neutral-meson mixing observables used in this work are also the basis of global flavor fits, which combine the measured mass differences and CP-violating quantities with CKM and lattice inputs. The recent CKMfitter and UTfit analyses test the consistency of the CKM unitarity triangle \cite{ValeSilva:2024jml,UTfit:2022hsi}, model-independent studies translate the $B_d$- and $B_s$-mixing measurements into constraints on NP contributions to the mixing amplitudes \cite{Charles:2020dfl}, and analyses of generic $\Delta F=2$ operators determine the scale at which new flavor-violating physics may appear \cite{UTfit:2007eik,Silvestrini:2018dos}. Our likelihood is not a replacement for these global analyses. It is a model-specific analysis in which the single coupling vector of the portal fixes the relative size and phases of the new contributions to $K$, $B_d$, and $B_s$ mixing.

In addition to meson mixing, the inclusive radiative decays $B\to X_s\gamma$ and $B\to X_d\gamma$ provide a complementary $\Delta F=1$ test. In the weak effective theory below the electroweak scale, new heavy particles contribute to these decays mainly through the electromagnetic and chromomagnetic dipole WCs $C_{7\gamma}$ and $C_{8G}$, whose matching and running can also be computed with public tools \cite{Aebischer:2018bkb,Straub:2018kue}. The SM prediction for $B\to X_s\gamma$ is known at next-to-next-to-leading order (NNLO) in QCD, and work continues on the charm-quark contribution and on power corrections \cite{Misiak:2015xwa,Misiak:2020vlo,Czaja:2023ren,Gunawardana:2019gep}. Since the portal generates dipole operators of the opposite chirality to the SM ones, its contribution does not interfere with the SM amplitude and enters the decay rate only quadratically, so that these decays give a weaker constraint than meson mixing. In this work, we compute the portal dipole WCs at leading-logarithmic (LL) accuracy and include the uncertainty of the SM prediction as a nuisance parameter. We also estimate the size of higher-order corrections using the two-loop anomalous dimensions of the dipole operators \cite{Misiak:1994zw}, but we keep this estimate separate from the fit, because the corresponding two-loop matching and decay matrix elements of the portal are not included. Our treatment of the radiative decays is therefore a leading-order rescaling of the SM prediction, and not a new NNLO calculation.

In this study, we compute the WCs of the electromagnetic and chromomagnetic dipole operators and of the $\Delta F=2$ operator generated by the portal for general down-type flavors, and evolve them to low energies, taking into account the change in the number of active quark flavors at the heavy-quark thresholds. We then construct a likelihood for seven observables: the $B_s$ and $B_d$ mass differences $\Delta M_s$ and $\Delta M_d$, the CP-violating quantities $\phi_s^{c\bar cs}$ and $\sin2\beta$ in $B_s$ and $B_d$ mixing, the kaon CP-violation parameter $|\epsilon_K|$, the branching ratio $\BR(B\to X_s\gamma)$, and the ratio $R_{d/s}^{\gamma}$ of the $b\to d\gamma$ and $b\to s\gamma$ rates. The SM predictions for these observables depend on theory inputs, namely the CKM parameters extracted from tree-level processes, the lattice-QCD matrix elements, and the short-distance (SD) factors of $\epsilon_K$ \cite{ValeSilva:2024jml,FlavourLatticeAveragingGroupFLAG:2024oxs,Brod:2019rzc}. Rather than propagating their uncertainties through a covariance evaluated at the SM point, we treat these inputs as nuisance parameters and fit them together with the portal parameters at every point, so that the theory uncertainties can adjust in the presence of NP. For $b\to d\gamma$, we use the BaBar measurement of $R_{d/s}^{\gamma}$, in which both decays are measured in the same range of hadronic mass, \changed{$0.5<m_X<2.0\,\mathrm{GeV}$}, so that the measurement is not compared with a fully inclusive prediction \cite{BaBar:2010vgu}. Since the portal angle and phases have no effect when the portal couplings vanish, the standard $\chi^2$ interpretation of the fit is not reliable near the SM, and we therefore calibrate it with pseudo-experiments at the 1 TeV benchmark. Finally, for the complex scalar, we determine the bound on the portal couplings to the light quarks as a function of the two new masses $M_D$ and $m_{\mathcal S}$.

We now give the structure of the rest of this paper. In section~\ref{sec:model}, we define the portal and the three flavor structures. In section~\ref{sec:matching}, we present the matching and the QCD evolution. In section~\ref{sec:observables}, we collect the flavor observables. In section~\ref{sec:likelihood}, we describe the likelihood and the numerical setup. In section~\ref{sec:results}, we present our results, and in section~\ref{sec:conclusions} we summarize our conclusions. Technical details are collected in the appendices.

\section{The right-handed $U(2)^3$ portal}
\label{sec:model}

\subsection{Field content and symmetries}

The new fields are a vector-like down quark $D$ and a neutral scalar $\mathcal S$,
\begin{equation}
D_{L,R}\sim(3,1,-1/3),\qquad
\mathcal S\sim(1,1,0),
\label{eq:new-fields}
\end{equation}
where the entries denote the representations under $SU(3)_c\times SU(2)_L\times U(1)_Y$. We consider two minimal choices for the scalar, namely a real scalar $\mathcal S=S_R=S_R^\dagger$ and a complex scalar $\mathcal S=\Phi\neq\Phi^\dagger$. Both $D$ and $\mathcal S$ are odd under a $Z_2$ parity, while all the SM fields are even, and we assume that $\mathcal S$ does not acquire a vacuum expectation value, $\langle\mathcal S\rangle=0$. The $Z_2$ parity allows the portal interaction between $\mathcal S$, $D$, and the right-handed down quarks, but it forbids the mixing of $D$ with the SM down quarks, both through the Yukawa term $\bar q_L H D_R$ and through the mass term $\bar D_L d_R$. The condition $\langle\mathcal S\rangle=0$ ensures that the portal interaction itself does not generate such a mixing. In addition, the $Z_2$ parity makes the lighter of the two new particles stable. For $m_{\mathcal S}<M_D$, the scalar is stable and is a dark-matter candidate, whereas for $M_D<m_{\mathcal S}$ the colored quark $D$ would be stable, which is excluded, as discussed in section~\ref{sec:nonflavor}. The quantum numbers of the new fields, together with those of the right-handed down quarks to which they couple, are summarized in table~\ref{tab:fields}.

\begin{table}[H]
\centering
\renewcommand{\arraystretch}{1.25}
\begin{tabular}{@{}lcccccc@{}}
\toprule
\rowcolor{tabhead}
\textbf{Field} & $SU(3)_c$ & $SU(2)_L$ & $U(1)_Y$ & $U(2)^3$ & $Z_2$
  & $U(1)_{\rm DS}$ \\
\midrule
$S_R=S_R^\dagger$ (real scalar) & $1$ & $1$ & $0$ & $(1,1,1)$ & $-$ & --- \\
\rowcolor{tabrow}
$\Phi\neq\Phi^\dagger$ (complex scalar) & $1$ & $1$ & $0$ & $(1,1,1)$ & $-$
  & $+1$ \\
$D_L$, $D_R$ & $3$ & $1$ & $-1/3$ & $(1,1,1)$ & $-$ & $+1$ \\
\rowcolor{tabrow}
$d_R^{(2)}=(d_R,s_R)^T$ & $3$ & $1$ & $-1/3$ & $(1,1,2)$ & $+$ & $0$ \\
$b_R$ & $3$ & $1$ & $-1/3$ & $(1,1,1)$ & $+$ & $0$ \\
\bottomrule
\end{tabular}
\caption{Quantum numbers of the new fields and of the right-handed down quarks to which the portal couples. The $U(2)^3$ column gives the representation under $U(2)_Q\times U(2)_u\times U(2)_d$ of eq.~\eqref{eq:flavor-group}. The $U(1)_{\rm DS}$ symmetry, introduced below, is imposed only for the complex scalar; the portal interaction $\Phi\,\bar D_L d_{Ri}$ is neutral under it.}
\label{tab:fields}
\end{table}

The $Z_2$ parity alone, however, is not sufficient to define the complex-scalar case, since it still allows the terms
\begin{equation}
\mu^2\Phi^2+\mathrm{h.c.},\qquad
\lambda_2'\,\Phi^2H^\dagger H+\mathrm{h.c.}
\label{eq:u1-violating-terms}
\end{equation}
These terms split $\Phi$ into two real scalars with different masses, so that $\Phi$ no longer carries a conserved charge. This conserved charge is precisely what distinguishes the complex case from the real one. To avoid this, we impose an exact global $U(1)_{\rm DS}$ symmetry under which $\Phi\to e^{i\alpha}\Phi$ and $D_{L,R}\to e^{i\alpha}D_{L,R}$, with all the SM fields neutral. The portal interaction $\Phi\,\bar D_Ld_{Ri}$ is invariant under this symmetry, whereas both terms in eq.~\eqref{eq:u1-violating-terms} carry charge $2$ and are therefore forbidden. A discrete $Z_4$ symmetry with $\Phi\to i\Phi$ and $D\to iD$ forbids the same terms and avoids an exact global continuous symmetry. With these assignments, the scalar potential can be written as
\begin{subequations}
\label{eq:scalar-potential}
\begin{align}
V_{\rm complex}&=m_{\mathcal S}^2|\Phi|^2+\lambda_\Phi|\Phi|^4
+\lambda_{\Phi H}|\Phi|^2H^\dagger H,\\
V_{\rm real}&=\tfrac12m_{\mathcal S}^2S_R^2+\tfrac{1}{4!}\lambda_SS_R^4
+\tfrac12\lambda_{SH}S_R^2H^\dagger H,
\end{align}
\end{subequations}
where the odd terms $S_R^3$ and $S_RH^\dagger H$ in the real case are forbidden by the $Z_2$ parity. After electroweak symmetry breaking, the Higgs-portal couplings $\lambda_{\Phi H}$ and $\lambda_{SH}$ only shift the scalar mass, and we denote the resulting physical mass by $m_{\mathcal S}$; they do not otherwise affect the flavor observables studied here. The condition $m_{\mathcal S}^2>0$ ensures that $\langle\mathcal S\rangle=0$, as assumed above.

The real and complex scalars can also be viewed as the two limits of a one-parameter family. If the $U(1)_{\rm DS}$ symmetry is broken by a nonzero $\mu^2$ in eq.~\eqref{eq:u1-violating-terms}, the complex scalar splits into two real scalars $S_{1,2}$ with masses $m_{1,2}^2=m_{\mathcal S}^2\mp2|\mu^2|$, where $\mu^2$ also includes the contribution $\lambda_2'v^2/2$ of the Higgs-portal term after electroweak symmetry breaking, with $v\simeq246\,\mathrm{GeV}$ the Higgs vacuum expectation value. The derivation of these masses, together with the resulting $\Delta F=2$ WC, is given in appendix~\ref{app:boxcancel}, where we show that the $\Delta F=2$ amplitude is then proportional to the product of the two scalar propagators. It therefore reduces to the complex-scalar result when the two states are degenerate, $\mu^2=0$, and it vanishes when one of them becomes very heavy, which corresponds to the single-real-scalar case. The size of the $U(1)_{\rm DS}$ breaking thus controls where the model lies between the two cases considered here.

We now turn to the flavor structure. Under the $U(2)^3$ symmetry of eq.~\eqref{eq:flavor-group}, the first two quark generations transform as
\begin{equation}
Q^{(2)}=\begin{pmatrix}Q_1\\Q_2\end{pmatrix}\sim(2,1,1),\qquad
u_R^{(2)}=\begin{pmatrix}u_R\\c_R\end{pmatrix}\sim(1,2,1),\qquad
d_R^{(2)}=\begin{pmatrix}d_R\\s_R\end{pmatrix}\sim(1,1,2),
\label{eq:u2-doublets}
\end{equation}
where $Q_1=(u_L,d_L)^T$ and $Q_2=(c_L,s_L)^T$ are the $SU(2)_L$ doublets of the first two generations. The third generation, $Q_3=(t_L,b_L)^T$, $t_R$, and $b_R$, as well as $D$ and $\mathcal S$, are singlets under $U(2)^3$, see also table~\ref{tab:fields}. In minimal $U(2)^3$, the observed quark masses and CKM mixing are reproduced by the spurions
\begin{equation}
\Delta Y_u\sim(2,\bar 2,1),\qquad
\Delta Y_d\sim(2,1,\bar 2),\qquad
V\sim(2,1,1).
\label{eq:minimal-spurions}
\end{equation}
A coupling of the portal to $d_R^{(2)}$ requires a spurion transforming as $(1,1,2)$. The spurion $V$ acts in the left-handed $U(2)_Q$ space and therefore cannot provide it by itself. In generic $U(2)^3$, this is an additional right-handed spurion \cite{Barbieri:2012bh},
\begin{equation}
V_d\sim(1,1,2),
\label{eq:vd-spurion}
\end{equation}
whereas in minimal $U(2)^3$ it must be built from the spurions of eq.~\eqref{eq:minimal-spurions}, as discussed in section~\ref{sec:portal-params}.

\subsection{Portal interaction and physical parameters}
\label{sec:portal-params}

At first order in the spurion $V_d$, the portal interaction is
\begin{equation}
\cL_{\rm portal}=-\mathcal S\,\bar D_L
\left(\lambda_b b_R+\lambda_V V_d^\dagger d_R^{(2)}\right)
+\mathrm{h.c.},
\label{eq:portal-spurion}
\end{equation}
where $\lambda_b$ and $\lambda_V$ are coupling constants. Eq.~\eqref{eq:portal-spurion} follows from three requirements. Since a scalar couples fermions of opposite chirality, gauge invariance allows the coupling $\mathcal S\,\bar D_Ld_{Ri}$, for which $(1,1,0)\otimes(\bar3,1,+1/3)\otimes(3,1,-1/3)$ contains a singlet, but it forbids $\mathcal S\,\bar D_RQ_{Li}$, because $Q_L$ is an $SU(2)_L$ doublet. The portal therefore couples only to the right-handed down quarks, $-\mathcal S\,\bar D_L\lambda_i d_{Ri}+{\rm h.c.}$  Under $U(2)^3$, $d_R^{(2)}\to U_d\,d_R^{(2)}$ while $b_R$ is a singlet, so that the coupling to $b_R$ is allowed directly, whereas the coupling to $d_R^{(2)}$ requires the spurion: with $V_d\to U_dV_d$, the combination $V_d^\dagger d_R^{(2)}$ is invariant. At first order in $V_d$, this gives eq.~\eqref{eq:portal-spurion}, and in the down-quark mass basis
\begin{equation}
\lambda_d=\lambda_V\,(V_d)_1^*,\qquad \lambda_s=\lambda_V\,(V_d)_2^* .
\label{eq:lambda-spurion}
\end{equation}
Together with $\lambda_b$, these couplings form a single vector, which we parameterize as
\begin{equation}
\begin{pmatrix}\lambda_d\\ \lambda_s\\ \lambda_b\end{pmatrix}
=\lambda_b
\begin{pmatrix}
\epsR\sin\thetaR\,e^{i\phi_d}\\
\epsR\cos\thetaR\,e^{i\phi_s}\\
1
\end{pmatrix}.
\label{eq:coupling-parametrization}
\end{equation}
Since $D$ appears only in its kinetic and mass terms and in eq.~\eqref{eq:portal-spurion}, the rephasing $D_{L,R}\to e^{i\alpha}D_{L,R}$ changes only the couplings, $\lambda_i\to\lambda_i e^{-i\alpha}$. We use it to make $\lambda_b$ real and positive, $\lambda_b=|\lambda_b|$, so that $\phi_d$ and $\phi_s$ are the phases of $\lambda_d$ and $\lambda_s$ relative to $\lambda_b$. Here $\epsR$ is the size of the couplings to the light quarks relative to $\lambda_b$, and $\tan\thetaR=|\lambda_d|/|\lambda_s|$ is their ratio. The independent parameters are therefore
\begin{equation}
\left\{M_D,m_{\mathcal S},|\lambda_b|,\epsR,\thetaR,\phi_d,\phi_s\right\},
\qquad 0\leq\thetaR\leq\frac{\pi}{2},
\label{eq:model-parameters}
\end{equation}
together with the choice of a real or a complex scalar. The flavor observables considered here depend on the couplings only through the products
\begin{equation}
\Lambda_{ij}\equiv\lambda_i^*\lambda_j,
\label{eq:lambda-products-definition}
\end{equation}
which we give explicitly below.

The four parameters $(\epsR,\thetaR,\phi_d,\phi_s)$ can describe any complex pair $(\lambda_d,\lambda_s)$\changed{, and in the numerical analysis we restrict them to $\epsR\leq1$, that is, to $|\lambda_d|^2+|\lambda_s|^2\leq|\lambda_b|^2$}. Generic $U(2)^3$ therefore predicts no relation between $\lambda_d$ and $\lambda_s$, since the size, direction, and phases of $V_d$ are not fixed by the quark masses and the CKM matrix \cite{Barbieri:2012bh}. Likewise, the rank-one relation among the flavor-changing couplings derived below follows from the single coupling vector, and not from the flavor symmetry.

In minimal $U(2)^3$, by contrast, $V_d$ is not an independent spurion. As discussed in section~\ref{sec:model}, a coupling to $d_R^{(2)}$ requires an object transforming as $(1,1,2)$, and the lowest-order combination of the minimal spurions with these quantum numbers is
\begin{equation}
V_d^{\rm min}\propto\Delta Y_d^\dagger V,
\label{eq:minimal-vd}
\end{equation}
in which the $U(2)_Q$ indices of $\Delta Y_d^\dagger\sim(\bar2,1,2)$ and $V\sim(2,1,1)$ are contracted, leaving a $U(2)_d$ doublet. \changed{The same combination appears in appendix~A of ref.~\cite{Barbieri:2012uh}, where the operator $\bar b_R\gamma^\mu V^\dagger\Delta Y_d\,d_R$ is given as an example of flavor violation among the right-handed quarks that arises only at higher order in the spurions.}

The consequence follows in three steps. First, $U(2)_Q$ can be used to rotate the left-handed spurion, and matching to the observed CKM matrix in the down-quark mass basis fixes its direction to
\begin{equation}
V\propto\begin{pmatrix}V_{td}^*\\ V_{ts}^*\end{pmatrix},
\label{eq:v-direction}
\end{equation}
which gives the known minimal-$U(2)^3$ relation $\Lambda_{db}^{\rm LH}/\Lambda_{sb}^{\rm LH}=V_{td}^*/V_{ts}^*$ in the left-handed sector \cite{Barbieri:2012uh}. Second, in the down-quark mass basis the light block of the down-Yukawa spurion is diagonal, $\Delta Y_d=\mathrm{diag}(y_d,y_s)$, up to corrections of higher order in the spurions. Third, combining the two gives
\begin{equation}
\left(V_d^{\rm min}\right)_i\propto\left(y_d V_{td}^*,\;y_sV_{ts}^*\right)_i ,
\label{eq:minimal-components}
\end{equation}
so that, by eq.~\eqref{eq:lambda-spurion}, the portal couplings carry the Yukawa couplings in addition to the CKM factors. With $y_q\propto m_q$, this gives
\begin{equation}
\frac{\Lambda_{db}}{\Lambda_{sb}}
=\frac{\lambda_d^*}{\lambda_s^*}
=\pm\frac{m_d}{m_s}\frac{V_{td}^*}{V_{ts}^*},
\qquad \tan\thetaR\simeq0.011 ,
\label{eq:minimal-u2}
\end{equation}
where the sign is not physical, as discussed below. Since the ratio of quark masses does not depend on the renormalization scale, we use $m_s/m_d=19.9(3)$, obtained from the FLAG $N_f=2+1+1$ averages \cite{FlavourLatticeAveragingGroupFLAG:2024oxs} $m_s/m_{ud}=27.227(81)$ \cite{ExtendedTwistedMass:2021gbo,EuropeanTwistedMass:2014osg,Bazavov:2017lyh,FermilabLattice:2014tsy} and $m_u/m_d=0.465(24)$ \cite{Giusti:2017dmp,MILC:2018ddw}. Eq.~\eqref{eq:minimal-u2} holds at first order in the spurion expansion; higher insertions such as $\Delta Y_d^\dagger\Delta Y_u\Delta Y_u^\dagger V$ are suppressed by $y_c^2$ and are negligible here.

\changed{However, eq.~\eqref{eq:minimal-u2} fixes only the direction of $(\lambda_d,\lambda_s)$, and not its size. For a coefficient of order one in eq.~\eqref{eq:minimal-vd} and $|\lambda_V|\simeq|\lambda_b|$, the spurion counting gives $|V_d^{\rm min}|\simeq(m_s/m_b)|V_{ts}|$, and therefore}
\begin{equation}
{\color{changedcolor}
\epsR^{\rm min}\simeq\frac{m_s}{m_b}\,|V_{ts}|\approx7.5\times10^{-4},}
\label{eq:natural-epsilon}
\end{equation}
\changed{where we have used $m_b/m_s=53.94(12)$ \cite{FermilabLattice:2018est} and $|V_{ts}|$ at the central CKM inputs of section~\ref{sec:inputs}. Since the $\Delta F=2$ amplitudes scale as $\epsR^2$, the portal effects in $B$ mixing are then about $3\times10^{-5}$ times smaller than at the fitted values of section~\ref{sec:alignment}, and minimal $U(2)^3$ with coefficients of order one cannot be distinguished from the SM. In this work, we therefore treat the size $\epsR$ as a free parameter and test only the direction predicted by eq.~\eqref{eq:minimal-u2}.}

It is useful to compare the minimal structure with a simpler ansatz, in which the portal couplings follow the CKM factors without the Yukawa couplings,
\begin{equation}
\frac{\Lambda_{db}}{\Lambda_{sb}}=\pm\frac{V_{td}^*}{V_{ts}^*}
\qquad\Longleftrightarrow\qquad
h_d=h_s,
\qquad \tan\thetaR\simeq0.21 ,
\label{eq:ckm-alignment}
\end{equation}
where $h_q$ is the relative NP correction to $B_q$ mixing, defined in eq.~\eqref{eq:hq-definition}. We stress that eq.~\eqref{eq:ckm-alignment} is a phenomenological benchmark and does not follow from minimal $U(2)^3$. We include it for two reasons. First, it carries over to the right-handed sector the pattern of the left-handed sector, where minimal $U(2)^3$ predicts the same relative correction to $B_d$ and $B_s$ mixing \cite{Barbieri:2012uh}. Second, it gives $h_d=h_s$ exactly and therefore provides a clean point of comparison. Both hypotheses fix $\thetaR$ and the phase difference $\phi_d-\phi_s$, so that only $\epsR$ and one common phase remain free. In section~\ref{sec:alignment}, we test both hypotheses against the generic case.

The two signs in eqs.~\eqref{eq:minimal-u2} and \eqref{eq:ckm-alignment} are not physically distinct. The mixing amplitudes depend on $\Lambda_{ij}^2$, and, as we show in section~\ref{sec:observables}, the radiative rates depend on $|\Lambda_{qb}|^2$, because the portal dipole does not interfere with the SM one. The shift $\phi_s\to\phi_s+\pi$ changes the sign of $\lambda_s$, and therefore of $\Lambda_{sb}$ and $\Lambda_{ds}$, while $\Lambda_{db}$ is unchanged. It reverses the sign of $\Lambda_{db}/\Lambda_{sb}$ but leaves every observable considered here unchanged, so that the two signs describe the same physics.

In terms of the parameters of eq.~\eqref{eq:coupling-parametrization}, the three flavor products relevant for the present analysis are
\begin{subequations}
\label{eq:lambda-products}
\begin{align}
\Lambda_{sb}&=|\lambda_b|^2\epsR\cos\thetaR\,e^{-i\phi_s},
\label{eq:lambda-sb}\\
\Lambda_{db}&=|\lambda_b|^2\epsR\sin\thetaR\,e^{-i\phi_d},
\label{eq:lambda-db}\\
\Lambda_{ds}&=|\lambda_b|^2\epsR^2\sin\thetaR\cos\thetaR\,
e^{i(\phi_s-\phi_d)}.
\label{eq:lambda-ds}
\end{align}
\end{subequations}
They obey the exact relations
\begin{equation}
\frac{\Lambda_{db}}{\Lambda_{sb}}
=\tan\thetaR\,e^{-i(\phi_d-\phi_s)},\qquad
\Lambda_{ds}=\frac{\Lambda_{db}\Lambda_{sb}^*}{|\lambda_b|^2}.
\label{eq:exact-flavor-correlations}
\end{equation}
The $s\to d$ product is therefore not an independent parameter, but is fixed by the two $b$-sector products. As noted above, this follows only from the single coupling vector: the matrix $\Lambda_{ij}=\lambda_i^*\lambda_j$ has rank one, and no flavor symmetry is needed. The observables that probe each product are listed in table~\ref{tab:flavor-map}.

\begin{table}[tbp]
\centering
\begin{tabular}{@{}lll@{}}
\toprule
Flavor transition & Portal product &  Observables \\
\midrule
$b\to s$ & $\Lambda_{sb}=\lambda_s^*\lambda_b$
& $B\to X_s\gamma$, $\Delta M_s$, $\phi_s^{c\bar cs}$ \\
$b\to d$ & $\Lambda_{db}=\lambda_d^*\lambda_b$
& $B\to X_d\gamma$, $\Delta M_d$, $\sin2\beta$ \\
$s\to d$ & $\Lambda_{ds}=\lambda_d^*\lambda_s$
& $|\epsilon_K|$ \\
\bottomrule
\end{tabular}
\caption{Flavor transitions, the corresponding portal products, and the observables used in the fit. The kaon Hamiltonian convention used below contains $(\Lambda_{ds}^*)^2$.}
\label{tab:flavor-map}
\end{table}

For a complex scalar, the mixing WCs are quadratic in the products of eqs.~\eqref{eq:lambda-products}, so that their phases are doubled: $-2\phi_s$ in $B_s$ mixing, $-2\phi_d$ in $B_d$ mixing, and $2(\phi_d-\phi_s)$ in the kaon Hamiltonian convention adopted below. The radiative rates, in contrast, depend only on $|\Lambda_{qb}|^2$ and carry no information on the phases. The phases are therefore constrained only by the mixing observables, while the radiative rates constrain only the size of the couplings.

\section{Effective theory, matching, and QCD evolution}
\label{sec:matching}

\subsection{Radiative dipoles}

The portal contribution to the radiative transitions $b\to q\gamma$ with $q=d,s$ is described by the effective Hamiltonian
\begin{equation}
\cH_{\rm eff}^{\Delta F=1}=-\frac{4\GF}{\sqrt2}V_{tb}V_{tq}^*
\left(C_{7\gamma}^{\prime\,qb}\cO_{7\gamma}^{\prime\,qb}
+C_{8G}^{\prime\,qb}\cO_{8G}^{\prime\,qb}+\cdots\right),
\label{eq:heff-dipole}
\end{equation}
with
\begin{subequations}
\label{eq:dipole-operators}
\begin{align}
\cO_{7\gamma}^{\prime\,qb}&=\frac{e}{16\pi^2}m_b
(\bar q\sigma^{\mu\nu}P_Lb)F_{\mu\nu},\\
\cO_{8G}^{\prime\,qb}&=\frac{g_s}{16\pi^2}m_b
(\bar q\sigma^{\mu\nu}T^aP_Lb)G^a_{\mu\nu},
\end{align}
\end{subequations}
and $P_{L,R}=(1\mp\gamma_5)/2$. The unprimed operators $\cO_{7\gamma}$ and $\cO_{8G}$, with $P_R$ in place of $P_L$, are the ones generated in the SM. The portal generates only the primed operators, because the interaction in eq.~\eqref{eq:portal-spurion} contains a single chirality. The incoming $b$ quark enters through $\bar DP_Rb$ and is therefore right-handed. The outgoing quark leaves through $\bar qP_LD$, and since $\bar qP_L=\overline{P_Rq}$, it is also right-handed. The amplitude therefore has the form $\bar q_R\Gamma b_R$, which vanishes for $\Gamma=\sigma^{\mu\nu}$. A chirality flip on the external $b$ line, proportional to $m_b$, is needed, and this gives the structure $\bar q\sigma^{\mu\nu}P_Lb$, in agreement with the general results of ref.~\cite{Arnan:2019uhr}. To keep the notation compact, we write $\Cseven^{qb}$ and $\Ceight^{qb}$ for the primed WCs in the rest of this paper, while the SM WC is always denoted $C_{7\gamma}^{\SM,\rm eff}$.

As a consequence, up to corrections of order $m_q/m_b$, the primed and unprimed dipoles do not interfere in the inclusive rate, so that the portal enters $\BR(B\to X_q\gamma)$ quadratically rather than linearly in the coupling product, see section~\ref{sec:observables}.

Since the scalar $\mathcal S$ is electrically neutral and a color singlet, the photon and the gluon can be emitted only from the internal $D$ line, as shown in figure~\ref{fig:dipole-diagram}. This fixes the relation between the two dipole WCs given below. The one-loop matching is a special case of the general scalar-fermion results of refs.~\cite{Bobeth:1999ww,Arnan:2019uhr}, and at the matching scale $\mu_D\simeq M_D$ we obtain
\begin{subequations}
\label{eq:dipole-matching}
\begin{align}
\Cseven^{qb,\NP}(\mu_D)&=
-\frac{\Lambda_{qb}}
{4\sqrt2\GF V_{tb}V_{tq}^*M_D^2}\,Q_D F(x),\\
\Ceight^{qb,\NP}(\mu_D)&=
-\frac{\Lambda_{qb}}
{4\sqrt2\GF V_{tb}V_{tq}^*M_D^2}\,F(x),
\end{align}
\end{subequations}
where
\begin{equation}
x=\frac{m_{\mathcal S}^2}{M_D^2},\qquad Q_D=-\frac13,
\label{eq:x-and-charge}
\end{equation}
and
\begin{equation}
F(x)=\frac{1-6x+3x^2+2x^3-6x^2\ln x}{12(1-x)^4}.
\label{eq:dipole-loop}
\end{equation}
The derivation is given in appendix~\ref{app:dipole}.
\begin{figure}[tbp]
\centering
\begin{tikzpicture}[scale=1.0]
  \coordinate (bin) at (-2.6,0);
  \coordinate (v1)  at (-1.3,0);
  \coordinate (v2)  at ( 1.3,0);
  \coordinate (qout) at (2.6,0);
  \coordinate (apex) at (0,1.35);
  \draw[fermion] (bin) -- (v1);
  \draw[fermion] (v2) -- (qout);
  \draw[heavyfermion] (v1) -- (apex);
  \draw[heavyfermion] (apex) -- (v2);
  \draw[scalarline] (v1) to[bend right=38] (v2);
  \draw[boson] (apex) -- ++(0,1.15);
  \node[vertexdot] at (v1) {};
  \node[vertexdot] at (v2) {};
  \node[vertexdot] at (apex) {};
  \node[left]  at (bin)  {$b_R$};
  \node[right] at (qout) {$q_R$};
  \node[left=3pt]  at ($(v1)!0.78!(apex)$) {$D$};
  \node[right=3pt] at ($(v2)!0.78!(apex)$) {$D$};
  \node[below] at (0,-0.62) {$\mathcal S$};
  \node[right] at (0.12,2.2) {$\gamma,\,g$};
  \node[below left]  at (v1) {$\lambda_b$};
  \node[below right] at (v2) {$\lambda_q^*$};
\end{tikzpicture}
\caption{One-loop Feynman diagram for $b\to q\gamma$ and $b\to qg$ in the portal. The dashed line is the scalar $\mathcal S$, the double line is the vector-like quark $D$, and the photon or gluon is emitted from $D$.}
\label{fig:dipole-diagram}
\end{figure}
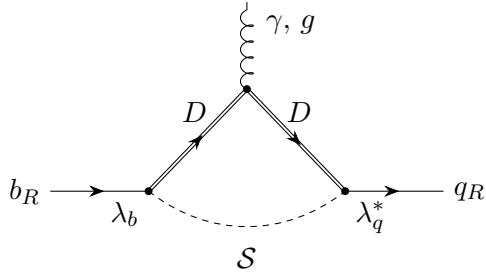

From eqs.~\eqref{eq:dipole-matching}, the two WCs at the matching scale are related by
\begin{equation}
\frac{\Cseven^{qb,\NP}(\mu_D)}{\Ceight^{qb,\NP}(\mu_D)}=Q_D=-\frac13,
\label{eq:dipole-ratio}
\end{equation}
that is, the electromagnetic WC equals the chromomagnetic one multiplied by the electric charge of $D$. This ratio is changed by the QCD evolution to low scales, discussed in section~\ref{sec:running}.

\subsection{Neutral-meson mixing}

\begin{figure}[tbp]
\centering
\begin{tikzpicture}[scale=0.95]
\begin{scope}
  \coordinate (tl) at (-1.15, 1.15);
  \coordinate (tr) at ( 1.15, 1.15);
  \coordinate (bl) at (-1.15,-1.15);
  \coordinate (br) at ( 1.15,-1.15);
  \draw[fermion] (-2.35, 1.15) -- (tl);
  \draw[fermion] (tr) -- ( 2.35, 1.15);
  \draw[fermion] (-2.35,-1.15) -- (bl);
  \draw[fermion] (br) -- ( 2.35,-1.15);
  \draw[heavyfermion] (tl) -- (tr);
  \draw[heavyfermion] (bl) -- (br);
  % A complex scalar joins a Phi vertex (incoming b) to a Phi^dagger vertex
  % (outgoing q): x1<->x4 and x2<->x3, as in appendix D.
  \draw[scalarline] (tl) -- (br);
  \draw[scalarline] (bl) -- (tr);
  \foreach \p in {(tl),(tr),(bl),(br)} \node[vertexdot] at \p {};
  \node[above left]  at (-2.35, 1.15) {$b$};
  \node[above right] at ( 2.35, 1.15) {$q$};
  \node[below left]  at (-2.35,-1.15) {$b$};
  \node[below right] at ( 2.35,-1.15) {$q$};
  \node[above] at (0, 1.15) {$D\;(k)$};
  \node[below] at (0,-1.15) {$D\;(k)$};
  \node[left=4pt]  at (-0.62, 0.62) {$\mathcal S$};
  \node[right=4pt] at ( 0.62, 0.62) {$\mathcal S$};
  \node at (0,-2.05) {direct};
\end{scope}
\begin{scope}[xshift=6.4cm]
  \coordinate (tl) at (-1.15, 1.15);
  \coordinate (tr) at ( 1.15, 1.15);
  \coordinate (bl) at (-1.15,-1.15);
  \coordinate (br) at ( 1.15,-1.15);
  \draw[fermion] (-2.35, 1.15) -- (tl);
  \draw[fermion] (tr) -- ( 2.35, 1.15);
  \draw[fermion] (-2.35,-1.15) -- (bl);
  \draw[fermion] (br) -- ( 2.35,-1.15);
  \draw[heavyfermion] (tl) -- (tr);
  \draw[heavyfermion] (bl) -- (br);
  % Only a real scalar can join the two incoming-b vertices, x1<->x2 and
  % x3<->x4; momentum conservation then reverses k on the lower D line.
  \draw[scalarline] (tl) -- (bl);
  \draw[scalarline] (tr) -- (br);
  \foreach \p in {(tl),(tr),(bl),(br)} \node[vertexdot] at \p {};
  \node[above left]  at (-2.35, 1.15) {$b$};
  \node[above right] at ( 2.35, 1.15) {$q$};
  \node[below left]  at (-2.35,-1.15) {$b$};
  \node[below right] at ( 2.35,-1.15) {$q$};
  \node[above] at (0, 1.15) {$D\;(k)$};
  \node[below] at (0,-1.15) {$D\;(-k)$};
  \node at (-1.55,0) {$\mathcal S$};
  \node at ( 1.55,0) {$\mathcal S$};
  \node at (0,-2.05) {crossed};
\end{scope}
\end{tikzpicture}
\caption{Direct and crossed box diagrams for $\Delta F=2$ transitions. A complex scalar allows only the direct diagram, while a real scalar allows both. The labels give the momenta of the two $D$ lines.}
\label{fig:box-diagrams}
\end{figure}
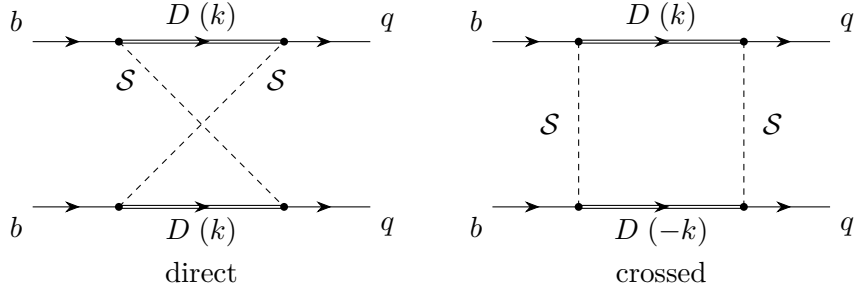

We now turn to the $\Delta F=2$ transitions. For $i\neq j$, the portal generates only the right-handed operator
\begin{equation}
\cO_{RR}^{ij}=(\bar d_i^\alpha\gamma_\mu P_Rd_j^\alpha)
(\bar d_i^\beta\gamma^\mu P_Rd_j^\beta),
\label{eq:orr-operator}
\end{equation}
with $\cH_{\rm eff}^{\Delta F=2}=\CRR^{ij}\cO_{RR}^{ij}+{\rm h.c.}$  For a complex scalar, only the direct box of figure~\ref{fig:box-diagrams} contributes, and it gives
\begin{equation}
\CRR^{ij}(\mu_D)=\frac{(\Lambda_{ij})^2}{128\pi^2M_D^2}
G_{\rm box}(x),
\label{eq:complex-box-coefficient}
\end{equation}
where
\begin{equation}
G_{\rm box}(x)=\frac{1-x^2+2x\ln x}{(1-x)^3}.
\label{eq:box-loop}
\end{equation}
The derivation is given in appendix~\ref{app:boxcancel}. For kaon mixing, we write the operator as $(\bar s\gamma_\mu P_Rd)^2$, so that its WC is proportional to $(\Lambda_{ds}^*)^2$. This conjugation fixes the sign of the portal contribution to $\epsilon_K$ in the CKM convention used in section~\ref{sec:results}.

For a real scalar, the scalar line has no orientation, and a crossed box accompanies the direct one, see figure~\ref{fig:box-diagrams}. The two diagrams have the same fermion contractions and the same denominators, but the second $D$ line carries the loop momentum $k$ in the direct box and $-k$ in the crossed box. Since only the $\slashed k$ part of each $D$ propagator survives the chiral projections, the two numerators are opposite, and
\begin{subequations}
\label{eq:real-box-cancellation}
\begin{align}
\CRR^{ij,\rm dir}(\mu_D)&=+
\frac{(\Lambda_{ij})^2}{128\pi^2M_D^2}G_{\rm box}(x),\\
\CRR^{ij,\rm cross}(\mu_D)&=-
\frac{(\Lambda_{ij})^2}{128\pi^2M_D^2}G_{\rm box}(x),\\
\CRR^{ij,\rm real}(\mu_D)&=0.
\end{align}
\end{subequations}
for all masses. This agrees with ref.~\cite{Arnan:2019uhr}, where the crossed box of a real scalar is the direct one with the opposite sign. A crossed contribution that does not cancel, proportional to the fermion masses, requires a Majorana fermion, which $D$ cannot be, since it carries color and electric charge.

The cancellation holds for $ij=sb$, $db$, and $sd$, since the loop is flavor blind. It is exact for a single real scalar, but it is not a general property of real scalars. If the complex scalar is split into two real states $S_{1,2}$ with masses $m_{1,2}$, the direct and crossed contributions combine into the product of the two propagators, see appendix~\ref{app:boxcancel}. The amplitude is then largest for degenerate states, where the complex-scalar result is recovered, and it vanishes when one of the two states decouples. Additional scalars or couplings to left-handed quarks also lift the cancellation.

\subsection{QCD evolution and higher-order effects}
\label{sec:running}

Between two scales $\mu_h>\mu_l$ with a fixed number $n_f$ of quark flavors, we define
\begin{equation}
\eta=\frac{\alpha_s(\mu_h)}{\alpha_s(\mu_l)}.
\label{eq:eta-definition}
\end{equation}
At LL accuracy, the dipole WCs evolve as \cite{Buchalla:1995vs,Chetyrkin:1996vx}
\begin{subequations}
\label{eq:dipole-running}
\begin{align}
\Cseven(\mu_l)&=\eta^{16/(33-2n_f)}\Cseven(\mu_h)
+\frac83\left[\eta^{14/(33-2n_f)}-
\eta^{16/(33-2n_f)}\right]\Ceight(\mu_h),\\
\Ceight(\mu_l)&=\eta^{14/(33-2n_f)}\Ceight(\mu_h).
\end{align}
\end{subequations}
and the $\Delta F=2$ WC as
\begin{equation}
\CRR(\mu_l)=\eta^{6/(33-2n_f)}\CRR(\mu_h).
\label{eq:vrr-running}
\end{equation}
We evolve from $\mu_D\simeq M_D$ with $n_f=6$ down to the top threshold at $m_t\simeq163\,\mathrm{GeV}$ \cite{ParticleDataGroup:2026mpi}, and with $n_f=5$ down to $\mu_b=4.8\,\mathrm{GeV}$, the $b$-quark pole mass \cite{Chetyrkin:1996vx,Salam:2026vys}. The kaon WC is evolved further through the bottom threshold at $m_b=4.18\,\mathrm{GeV}$ \cite{ParticleDataGroup:2024cfk} down to $\mu_K=3\,\mathrm{GeV}$ \cite{FlavourLatticeAveragingGroupFLAG:2024oxs}. At each threshold, $\alpha_s$ is matched continuously at one loop. These LL expressions are used in the fit.

To estimate the size of higher-order terms, we also evolve the NP WCs $\bm C=(C_{7\gamma},C_{8G})^T$ with the two-loop anomalous dimensions and the two-loop QCD beta function,
\begin{equation}
\frac{d\bm C}{d\ln\mu}=\left[
\frac{\alpha_s}{4\pi}\widehat\gamma^{(0)T}
+\left(\frac{\alpha_s}{4\pi}\right)^2\widehat\gamma^{(1)T}
\right]\bm C,
\label{eq:partial-nll-rge}
\end{equation}
where, with $N=3$, $C_F=4/3$, and the down-quark charge $Q_d=-1/3$,
\begin{subequations}
\label{eq:dipole-adm}
\begin{align}
\widehat\gamma^{(0)}&=
\begin{pmatrix}8C_F&0\\8C_FQ_d&16C_F-4N\end{pmatrix},\\
\widehat\gamma^{(1)}_{77}&=C_F\left(\frac{548}{9}N-16C_F-
\frac{56}{9}n_f\right),\qquad \widehat\gamma^{(1)}_{78}=0,\\
\widehat\gamma^{(1)}_{87}&=Q_dC_F\left(\frac{404}{9}N-32C_F-
\frac{56}{9}n_f\right),\\
\widehat\gamma^{(1)}_{88}&=-\frac{458}{9}-\frac{12}{N^2}
+\frac{214}{9}N^2+\frac{56}{9N}n_f-\frac{13}{9}n_fN.
\end{align}
\end{subequations}
For $n_f=5$, these give $\widehat\gamma_{77}^{(1)}=4688/27$, $\widehat\gamma_{87}^{(1)}=-2192/81$, and $\widehat\gamma_{88}^{(1)}=4063/27$, the values of ref.~\cite{Misiak:1994zw}. Equation~\eqref{eq:partial-nll-rge} is evolved across the top threshold in the same way as the LL evolution. This resums the two-loop self-mixing of the dipoles, but it is not a complete next-to-leading-logarithmic (NLL) prediction, which would also require the $\mathcal O(\alpha_s)$ matching of the portal and the next-to-leading-order (NLO) matrix elements \cite{Bobeth:1999ww,Buchalla:1995vs,Chetyrkin:1996vx}. We therefore do not use it in the fit. In section~\ref{sec:validation}, we show that it changes $\Cseven^{\NP}(\mu_b)$ by about $25\%$, but the observables by less than $10^{-6}$, because the portal dipole is very small compared with the SM one.

\section{Flavor observables}
\label{sec:observables}

\subsection{$B_q$ mixing and CP phases}

In this section, we collect the expressions for the observables used in the likelihood. For $B_q$ mixing, the SM dispersive amplitude is \cite{Buras:1990fn,Buchalla:1995vs,Lenz:2006hd}
\begin{equation}
M_{12}^{q,\SM}=\frac{\GF^2M_W^2}{12\pi^2}
M_{B_q}\eta_B S_0(x_t)
\left(f_{B_q}\sqrt{\widehat B_{B_q}}\right)^2
(V_{tq}^*V_{tb})^2,
\qquad q=d,s,
\label{eq:sm-b-mixing}
\end{equation}
where $S_0(x_t)$ is the Inami-Lim function \cite{Inami:1980fz}, $x_t=\overline m_t^2/M_W^2$, and $\Delta M_q^{\SM}=2|M_{12}^{q,\SM}|$. By parity, the left-handed SM operator and the right-handed portal operator have the same hadronic matrix element. The relative NP correction is therefore
\begin{equation}
h_q=\frac{\widehat C_{RR}^{qb}}{C_{LL}^{q,\SM}},
\qquad
C_{LL}^{q,\SM}=\frac{\GF^2M_W^2}{4\pi^2}
\eta_BS_0(x_t)(V_{tq}^*V_{tb})^2.
\label{eq:hq-definition}
\end{equation}
Here $\eta_B$ is defined as in ref.~\cite{Buras:1990fn}, so that $C_{LL}^{q,\SM}$ multiplies the renormalization-group-invariant (RGI) bag parameter $\widehat B_{B_q}$. The portal WC is brought to the same normalization through $\widehat C_{RR}^{qb}=C_{RR}^{qb}(\mu_b)\,[\alpha_s(\mu_b)]^{6/23}$, so that
both WCs in $h_q$ are RGI.\footnote{%
Equation~(6.18) of ref.~\cite{Salam:2026vys} compared $C_{RR}$ at $\mu_b$ with the RGI SM WC. The consistent RGI normalization used here reduces $h_q$ by the factor $[\alpha_s(\mu_b)]^{6/23}=0.66$.}  The kaon amplitude in eq.~\eqref{eq:np-kaon-m12} is written with $B_K^{\overline{\rm MS}}(\mu_K)$ instead, so that its WC needs no such conversion. The observables are then \cite{Lenz:2006hd}
\begin{subequations}
\label{eq:b-mixing-observables}
\begin{align}
\Delta M_q&=\Delta M_q^{\SM}|1+h_q|,\\
\phi_s^{c\bar cs}&=\phi_s^{c\bar cs,\SM}+\arg(1+h_s),\\
{\color{changedcolor}2\beta_{\rm eff}}&{\color{changedcolor}=2\beta+\arg(1+h_d)}.
\end{align}
\end{subequations}
The correlation between the $B_s$ and $B_d$ matrix elements is included in their joint nuisance prior, see section~\ref{sec:inputs}. Reviews of the SM mixing formalism and of its theoretical precision are given in refs.~\cite{Lenz:2006hd,DiLuzio:2019jyq}.

\subsection{Manifest-unitarity formulation of $\epsilon_K$}

Let
\begin{equation}
\lambda_i=V_{is}^*V_{id},\qquad i=u,c,t.
\label{eq:lambda-ckm-kaon}
\end{equation}
Instead of the conventional charm-top ($c$-$t$) form, we use the manifest-unitarity $u$-$t$ form of the SD amplitude \cite{Brod:2019rzc}. The part that contributes to the imaginary amplitude is
\begin{equation}
\lambda_t^2\eta_{tt}\overline S_t
+2\lambda_u\lambda_t\eta_{ut}\overline S_{ut},
\label{eq:epsilon-ut-amplitude}
\end{equation}
where
\begin{equation}
\overline S_{ut}=S_0(x_c)-S_0(x_c,x_t),\qquad
\overline S_t=S_0(x_t)+S_0(x_c)-2S_0(x_c,x_t).
\label{eq:sbar-definitions}
\end{equation}
In this form, the imaginary part does not depend on the charm-charm factor $\eta_{cc}$, whose perturbative expansion converges poorly, and $\eta_{ut}$ is known with a ten times smaller uncertainty than the charm-top factor $\eta_{ct}=0.496(47)$ of the $c$-$t$ form \cite{Brod:2010mj,Brod:2019rzc}. We use $\eta_{tt}=0.55(1\pm4.2\%)$ and $\eta_{ut}=0.402(5)$ \cite{Brod:2019rzc}, with $\eta_{ut}$ rescaled by $1.005$ for the electroweak corrections \cite{Brod:2022har}.

By parity, the portal operator has the same matrix element as the left-handed one, $\langle\bar K^0|\cO_{RR}^{sd}|K^0\rangle=\frac23M_K^2F_K^2 B_K^{\overline{\rm MS}}(\mu)$ \cite{Buchalla:1995vs}. The portal contribution at $\mu_K=3\,\mathrm{GeV}$ is therefore
\begin{equation}
M_{12}^{K,\NP}\equiv\frac{\langle\bar K^0|\cH_{\rm eff}^{\Delta F=2,\NP}|K^0\rangle}
{2M_K}=C_{RR}^{sd}(\mu_K)
\frac{M_KF_K^2B_K^{\overline{\rm MS}}(\mu_K)}{3}.
\label{eq:np-kaon-m12}
\end{equation}
With $\epsilon_K=\kappa_\epsilon e^{i\phi_\epsilon}\operatorname{Im}M_{12}/ (\sqrt2\Delta M_K)$ of ref.~\cite{Buras:2010pza}, in which $M_{12}$ is the complex conjugate of eq.~\eqref{eq:np-kaon-m12}, we obtain
\begin{equation}
\epsilon_K^{\NP}=-\frac{\kappa_\epsilon}{\sqrt2\Delta M_K}
\operatorname{Im}M_{12}^{K,\NP}.
\label{eq:np-epsilon-k}
\end{equation}
where $\kappa_\epsilon=0.94(2)$ accounts for long-distance contributions and for $\phi_\epsilon\neq45^\circ$ \cite{Buras:2010pza}. The derivation of the sign is given in appendix~\ref{app:conventions}. We use $|\epsilon_K|$ in the likelihood. We do not use $\Delta M_K$ as a constraint, because of its large long-distance contribution.

\subsection{Radiative observables}

The inclusive $B\to X_s\gamma$ average is compared with the SM prediction at $E_\gamma>1.6\,\mathrm{GeV}$ \cite{HeavyFlavorAveragingGroupHFLAV:2024ctg,Misiak:2020vlo}. Because the portal generates the opposite-chirality dipole $\cO_{7\gamma}^{\prime\,qb}$, its amplitude does not interfere with the SM one up to terms of order $m_q/m_b$, and the rate is a sum of squares,
\begin{equation}
\BR_q=\BR_q^{\SM}
\left(1+\left|\frac{C_{7\gamma}^{qb,\NP}(\mu_b)}
{C_{7\gamma}^{\SM,\rm eff}(\mu_b)}\right|^2\right),
\label{eq:radiative-rescaling}
\end{equation}
where $C_{7\gamma}^{\SM,\rm eff}(\mu_b)=-0.30$ is the leading-order SM value \cite{Buras:1998raa}. The quadratic dependence weakens the radiative constraint strongly compared with a portal of the same chirality as the SM, for which the leading effect would be linear. For $b\to d\gamma$, we use the BaBar ratio
\begin{equation}
R_{d/s}^{\gamma}=\frac{\BR(b\to d\gamma)}{\BR(b\to s\gamma)}
=0.040\pm0.009_{\rm stat}\pm0.010_{\rm syst},
\label{eq:radiative-ratio-experiment}
\end{equation}
measured in the same hadronic-mass window $0.5<m_X<2.0\,\mathrm{GeV}$ for both modes \cite{BaBar:2010vgu}. This avoids comparing a partial experimental $b\to d\gamma$ rate with a fully inclusive theory prediction. Since the $b\to s\gamma$ and $b\to d\gamma$ spectra are expected to be almost identical \cite{BaBar:2010vgu}, the extrapolation to the full mass range cancels in the ratio. We use the BaBar conversion to $|V_{td}/V_{ts}|$ in the form
\begin{equation}
R_{d/s}^{\gamma}=\kappa_{d/s}^{\gamma}
\left|\frac{V_{td}}{V_{ts}}\right|^2
\frac{1+\left|C_{7\gamma}^{db,\NP}/C_{7\gamma}^{\SM,\rm eff}\right|^2}
{1+\left|C_{7\gamma}^{sb,\NP}/C_{7\gamma}^{\SM,\rm eff}\right|^2}.
\label{eq:radiative-ratio-profile}
\end{equation}
BaBar converts $R_{d/s}^{\gamma}=0.040$ into $|V_{td}/V_{ts}|=0.199$, with a theory uncertainty of $0.002$ \cite{BaBar:2010vgu}, which gives
\begin{equation}
\kappa_{d/s}^{\gamma}=\frac{0.040}{0.199^2}=1.0101\pm0.0203.
\label{eq:kappa-radiative-ratio}
\end{equation}
Equation~\eqref{eq:radiative-ratio-profile} applies here because the portal changes only the normalization of the dipole, and not the shape of the photon spectrum. The portal WCs are evolved at LL accuracy, as in eq.~\eqref{eq:dipole-running}.
\subsection{Semileptonic transitions}
\label{sec:semileptonic}

The same couplings could in principle contribute to $b\to s\ell^+\ell^-$ and $b\to s\nu\bar\nu$. Applying the general scalar-fermion results of ref.~\cite{Arnan:2019uhr} to the present field content shows that almost all of these contributions vanish.

First, the box diagrams vanish identically. A semileptonic box requires either the scalar or the fermion in the loop to couple to both quarks and leptons, whereas $\mathcal S$ and $D$ couple only to $d_{Ri}$. This removes the box contributions to $C_{9,10}^{(\prime)}$, $C_{S,P}^{(\prime)}$, $C_T$, and $C_{L,R}^{\nu}$.

Second, the $Z$ penguin vanishes at $q^2=0$. Without electroweak symmetry breaking in the NP sector, the self-energy and vertex contributions cancel \cite{Arnan:2019uhr}. In the present model $\langle\mathcal S\rangle=0$, and the $Z_2$ parity forbids the mixing term $\bar q_LHD_R$, so there is no such breaking. This removes the $Z$-penguin contributions to $C_{9,10}^{(\prime)}$ and to $b\to s\nu\bar\nu$.

The only remaining contribution is the off-shell photon penguin. Since the portal couples only to right-handed quarks, it generates only
\begin{equation}
\cO_9^{\prime\,qb}=\frac{e^2}{16\pi^2}(\bar q\gamma^\mu P_Rb)
(\bar\ell\gamma_\mu\ell),
\end{equation}
with $\cH_{\rm eff}\supset-\frac{4\GF}{\sqrt2}V_{tb}V_{tq}^* C_9^{\prime\,qb}\cO_9^{\prime\,qb}$. With the photon emitted from $D$ only, we find
\begin{equation}
C_9^{\prime\,qb}=-\frac{\Lambda_{qb}}
{2\sqrt2\,\GF V_{tb}V_{tq}^*M_D^2}\,Q_D\,G_9(x),
\label{eq:c9-prime}
\end{equation}
with $x=m_{\mathcal S}^2/M_D^2$ as in eq.~\eqref{eq:x-and-charge}, and $G_9$ given in eq.~\eqref{eq:penguin-loop}, with $G_9(1)=1/8$. The loop function agrees with refs.~\cite{Hisano:1995cp,Arnan:2019uhr}, and the derivation is given in appendix~\ref{app:dipole}. At the benchmark $M_D=m_{\mathcal S}=1\TeV$ with $|\Lambda_{sb}|=1$, this gives $|C_9'|=3.1\times10^{-2}$, below $1\%$ of the SM value $C_9(\mu_b)=4.05$ \cite{Blake:2016olu}. With the $\Delta F=2$ bound $|\Lambda_{sb}|\leq\epsR^{95}|\lambda_b|^2=\changed{0.227}$ of section~\ref{sec:epsilon-profile}, $|C_9'|$ is below $10^{-2}$.

Consequently, $B_s\to\mu^+\mu^-$ and $B\to K^{(*)}\nu\bar\nu$ are not affected at one loop in this model, and $b\to s\ell^+\ell^-$ gives no useful constraint. These channels are otherwise among the most sensitive probes of right-handed $b\to s$ currents \cite{Belle-II:2023esi,Altmannshofer:2021qrr,Aliev:2001in,Aliev:2001fc}, but the portal cannot reach them. The flavor reach of the portal therefore rests on the $\Delta F=2$ observables, with the radiative rates entering only quadratically. This is why the likelihood below is restricted to these observables. It also sharpens the real-scalar case: for a single real scalar, the boxes cancel and the dipoles do not interfere, so that the leading flavor observables are effectively blind to it.

\section{Likelihood and numerical setup}
\label{sec:likelihood}

In this section, we describe the likelihood, the input parameters, and the numerical setup. The observables used in the fit are
\begin{equation}
\bm O=(\Delta M_s,\Delta M_d,\phi_s^{c\bar cs},{\color{changedcolor}2\beta_{\rm eff}},|\epsilon_K|,
\BR(B\to X_s\gamma),R_{d/s}^{\gamma}).
\label{eq:observable-vector}
\end{equation}
The neutral-meson inputs are HFLAV averages: $\Delta M_s=17.766(6)\, \mathrm{ps}^{-1}$, $\Delta M_d=0.5069(19)\,\mathrm{ps}^{-1}$, and $\phi_s^{c\bar cs}=-0.052(13)$ are \changed{taken from} ref.~\cite{HeavyFlavorAveragingGroupHFLAV:2024ctg}, and $\sin2\beta=0.710(11)$ is the all-charmonium average of the Summer 2025 \changed{online} update \cite{HFLAV:Summer2025}\changed{, obtained with the procedure of ref.~\cite{HFLAV:2022esi} from the measurements of BaBar, Belle, Belle~II, and LHCb \cite{BaBar:2009byl,Belle:2012paq,Belle-II:2024lwr,LHCb:2023zcp}}. \changed{Since $\sin2\beta$ alone does not distinguish $2\beta_{\rm eff}$ from $\pi-2\beta_{\rm eff}$, we fit the angle itself, $2\beta_{\rm eff}=\arcsin(0.710)=(45.24\pm0.90)^\circ$, on the branch with $\cos2\beta_{\rm eff}>0$. The other branch is excluded at $7.3$ standard deviations by the joint BaBar and Belle analysis of $B^0\to D^{(*)}h^0$ decays \cite{BaBar:2018agf}. Without this restriction, a second solution with $\cos2\beta_{\rm eff}\simeq-0.70$ and $|h_d|\simeq1.4$ would reproduce the measured $\sin2\beta$. The same likelihood is used for the pseudo-experiments and the mass plane in section~\ref{sec:results}.} The $\Delta M_s$ average is dominated by LHCb \cite{LHCb:2021moh}, and the $\phi_s^{c\bar cs}$ average by LHCb and \changed{CMS \cite{LHCb:2023sim,CMS:2024znt}, with a smaller contribution from ATLAS \cite{ATLAS:2020lbz}}. We use the averages rather than the individual results, so that correlated systematic uncertainties are treated as in HFLAV. The two sets of averages come from different HFLAV subgroups and are released at different times. For the kaon system, we use $|\epsilon_K|=2.228(11)\times10^{-3}$ \cite{ParticleDataGroup:2024cfk}. For the radiative decays, we use $\BR(B\to X_s\gamma)=3.49(19)\times10^{-4}$ for $E_\gamma>1.6\,\mathrm{GeV}$ \cite{HeavyFlavorAveragingGroupHFLAV:2024ctg}, with the SM prediction $3.40(17)\times10^{-4}$ \cite{Misiak:2020vlo}, and $R_{d/s}^{\gamma}$ of eq.~\eqref{eq:radiative-ratio-experiment}. A full covariance of these seven observables is not publicly available, so we treat their experimental uncertainties as uncorrelated.

Instead of a covariance fixed at the SM point, we keep the theory inputs as nuisance parameters that can adjust in the fit. With their standard-normal coordinates $\bm z$, we minimize
\begin{equation}
\chi^2(\bm p,\bm z)=
\sum_{a=1}^{7}
\frac{[O_a^{\rm th}(\bm p,\bm z)-O_a^{\rm exp}]^2}
{\sigma_{a,\rm exp}^2}
+\bm z^T\bm z,
\label{eq:chi2-definition}
\end{equation}
where $\bm p=(\epsR,\thetaR,\phi_d,\phi_s)$ are the portal parameters for fixed masses and fixed $|\lambda_b|$. All nuisance coordinates are re-optimized at every portal point. The fifteen coordinates are the four CKM parameters, the two correlated neutral-$B$ matrix elements, seven kaon and SD inputs, the SM $B\to X_s\gamma$ normalization, and $\kappa_{d/s}^{\gamma}$ of eq.~\eqref{eq:kappa-radiative-ratio}.

\subsection{Input parameters}
\label{sec:inputs}

\changed{For the CKM matrix, we use the tree-level inputs of the CKMfitter analysis of ref.~\cite{ValeSilva:2024jml}, which are not affected by the portal,}
\begin{equation}
{\color{changedcolor}
\begin{gathered}
|V_{cb}|=(41.22\pm0.24\pm0.37)\times10^{-3},\qquad
|V_{ub}|=(3.86\pm0.07\pm0.12)\times10^{-3},\\
\gamma=\left(65.9^{+3.3}_{-3.5}\right)^\circ,
\end{gathered}}
\label{eq:ckm-inputs}
\end{equation}
\changed{together with $\lambda=0.22498^{+0.00022}_{-0.00021}$ from the tree-level fit of the same reference. We add the statistical and theoretical errors in quadrature and treat the four inputs as independent nuisance parameters, with split-normal priors for the asymmetric intervals of $\gamma$ and $\lambda$. The Wolfenstein parameters $A$, $\bar\rho$, and $\bar\eta$ are then derived from these inputs exactly, so that their correlations are included automatically. At the central values, $A=0.814$, $\bar\rho=0.166$, and $\bar\eta=0.370$. Since CKMfitter treats the theoretical uncertainties in the Rfit scheme, the Gaussian treatment is an approximation, and in section~\ref{sec:alignment} we compare it with alternative treatments.} It is worth mentioning that an exactly unitary CKM matrix is reconstructed algebraically from these parameters, so that no truncated Wolfenstein expansion enters the observables.

For neutral-$B$ mixing, we use the $N_f=2+1+1$ FLAG values \cite{FlavourLatticeAveragingGroupFLAG:2024oxs}
\begin{equation}
f_{B_d}\sqrt{\widehat B_{B_d}}=210.6(5.5)\,\mathrm{MeV},\qquad
f_{B_s}\sqrt{\widehat B_{B_s}}=256.1(5.7)\,\mathrm{MeV},\qquad
\xi=1.216(16),
\label{eq:flag-b-inputs}
\end{equation}
which are the results of the HPQCD calculation of ref.~\cite{Dowdall:2019bea}. For the kaon, we use the $N_f=2+1$ FLAG average $\widehat B_K=0.7533(91)$ \cite{FlavourLatticeAveragingGroupFLAG:2024oxs} of the lattice results in refs.~\cite{RBC:2014ntl,BMW:2011zrh,Laiho:2011np,SWME:2015oos,Boyle:2024gge}. The $N_f=2+1$ value is used because the $\epsilon_K$ Hamiltonian is defined in the three-flavor theory below the charm threshold. Writing $X_s=f_{B_s}\sqrt{\widehat B_{B_s}}$ and $X_d=f_{B_d}\sqrt{\widehat B_{B_d}}$, the uncertainty on $\xi=X_s/X_d$ determines the linearized correlation
\begin{equation}
\rho_{sd}=
\frac{(\sigma_s/X_s)^2+(\sigma_d/X_d)^2-(\sigma_\xi/\xi)^2}
{2(\sigma_s/X_s)(\sigma_d/X_d)}=0.864.
\label{eq:lattice-correlation}
\end{equation}
This correlation is imposed through a two-dimensional Gaussian nuisance prior, constructed from the Cholesky factor of the implied covariance. The remaining SD inputs are $\eta_B=0.55(1)$ \cite{Buras:1990fn,Buchalla:1995vs}, $\overline m_c(\overline m_c)=1.2729(45)\,\mathrm{GeV}$ \cite{ParticleDataGroup:2026mpi}, and $\overline m_t(\overline m_t)=163.48(86)\,\mathrm{GeV}$ \cite{Brod:2019rzc}, the top mass used for $\eta_{tt}$ and $\eta_{ut}$ in ref.~\cite{Brod:2019rzc}. Together with $\widehat B_K$, $\eta_{tt}$, $\eta_{ut}$, and $\kappa_\epsilon$ of section~\ref{sec:observables}, they are profiled independently.

\subsection{Benchmark, scan ranges and calibration}
\label{sec:scansetup}

The benchmark fixes
\begin{equation}
M_D\simeq m_{\mathcal S}=1\TeV,\qquad |\lambda_b|=1,
\label{eq:mass-benchmark}
\end{equation}
and varies
\begin{equation}
10^{-6}\leq\epsR\leq1,\qquad
0\leq\thetaR\leq\frac{\pi}{2},\qquad
-\pi\leq\phi_d,\phi_s\leq\pi.
\label{eq:scan-ranges}
\end{equation}
\changed{The upper limit $\epsR\leq1$ is a choice of the scan range. It keeps the light-quark couplings below $|\lambda_b|$, as expected for a small spurion $V_d$ with $|\lambda_V|\simeq|\lambda_b|$, but it is not derived from the spurion counting, which gives a much smaller value in minimal $U(2)^3$, see eq.~\eqref{eq:natural-epsilon}. It is not a perturbativity bound either.} Since $\Lambda_{sb},\Lambda_{db}\propto |\lambda_b|^2$ at fixed $\epsR$, the 95\% CL upper limits $\epsR^{95}$ derived below scale approximately as $|\lambda_b|^{-2}$.

The equal mass benchmark is the limit of a slightly split spectrum. For $M_D=m_{\mathcal S}$, the decay $D\to q\,\mathcal S$ has no phase space, and $D$ would be a stable colored relic. The decay $D\to b\,\mathcal S$ requires $M_D>m_{\mathcal S}+m_b$, that is, a relative splitting $(M_D-m_{\mathcal S})/M_D>m_b/M_D\simeq5\times10^{-3}$ at $M_D=1\TeV$, with the $b$-quark pole mass $m_b=4.78\,\mathrm{GeV}$ \cite{ParticleDataGroup:2024cfk}, for example $m_{\mathcal S}=995\,\mathrm{GeV}$ for $M_D=1\TeV$. Since the loop functions $F$ and $G_{\rm box}$ are smooth at degeneracy, such a splitting changes the WCs by less than 1\%, and we therefore set $M_D=m_{\mathcal S}=1\TeV$ in the numerical analysis.

The minimization uses several starting values for the phases. For the profile in $\epsR$, we minimize over $\thetaR$, $\phi_d$, $\phi_s$, and all fifteen nuisance coordinates. The asymptotic 95\% CL upper limit $\epsR^{95}$ is the value at which
\begin{equation}
\Delta\chi^2(\epsR)=3.841.
\label{eq:profile-threshold}
\end{equation}
the 95\% quantile of a $\chi^2$ distribution with one degree of freedom, which we denote by $\chi^2_1$. This threshold cannot be taken for granted here. Because the portal phases are not identified at $\epsR=0$, the null point is not regular, and the conditions of Wilks' theorem \cite{Wilks:1938dza} are not satisfied there. Two effects are at work. First, $\epsR$ is bounded from below at the null, which gives a mixture distribution rather than a $\chi^2$ \cite{Chernoff:1954eli}. Second, the portal phases enter only for $\epsR>0$, so that they are nuisance parameters present only under the alternative \cite{Davies:1987zz}. The standard asymptotic formulae \cite{Cowan:2010js} account for the first effect, but not for the second. We therefore calibrate the profile statistic with 1000 pseudo-experiments at each of three truth points: the SM boundary, the observed best fit, and the asymptotic upper limit. Each pseudo-experiment fluctuates the seven observables and the fifteen nuisance constraints, and the fits with fixed and free $\epsR$ are then repeated.

The mass dependence of the upper limit is evaluated on a $7\times7$ logarithmic grid over
\begin{equation}
0.5\TeV\leq M_D,m_{\mathcal S}\leq5\TeV.
\label{eq:mass-plane-range}
\end{equation}
At every grid point, the four flavor parameters and all nuisances are profiled, and each upper limit is refined with several phase starts. Where an upper limit exists, $\Delta\chi^2$ at the quoted limit differs from $3.841$ by less than \changed{$0.008$}. \changed{At one grid point, the profile does not reach the threshold below $\epsR=1$}, see section~\ref{sec:mass-plane}.

\section{Profiled flavor results}
\label{sec:results}

\subsection{Validation of the matching and running}
\label{sec:validation}

Before presenting the fit, we give the WCs at the benchmark point of ref.~\cite{Salam:2026vys}, $M_D=m_{\mathcal S}=1\TeV$, $|\lambda_s^*\lambda_b|=1$, and $|V_{tb}V_{ts}^*|=0.040$. At the matching scale, they are
\begin{equation}
|C_{7\gamma}^{\NP}(\mu_D)|=5.2625\times10^{-3},\qquad
|C_{8G}^{\NP}(\mu_D)|=1.5788\times10^{-2},
\label{eq:matching-validation}
\end{equation}
and the evolution with quark thresholds to $\mu_b$ gives
\begin{equation}
|C_{7\gamma}^{\NP,\rm eff}(\mu_b)|=1.1263\times10^{-3}.
\label{eq:threshold-validation}
\end{equation}
In the fit, the WCs are normalized with the unitary CKM matrix at each nuisance point, for which $|V_{tb}V_{ts}^*|=\changed{0.0405}$ at the central inputs.

As a further check, we compare the $\Delta F=2$ normalization with the independent code {flavio} \cite{Straub:2018kue,Aebischer:2018bkb}. Since $\Delta M_s\propto|C_{LL}^{s,\SM}+C_{RR}^{sb}|$, the response of $\Delta M_s$ to a small $C_{RR}^{sb}$ in {flavio} gives the SM WC $C_{LL}^{s,\SM}$ used by that code, where it is called $C_{VLL}$. At $\mu=4.2\,\mathrm{GeV}$, it agrees with our value in eq.~\eqref{eq:hq-definition} to \changed{$4.3\%$}, consistent with the different CKM and SD inputs of the two codes. This comparison fixes the magnitude only; the relative sign of the SM and portal amplitudes follows from the conventions of appendix~\ref{app:conventions}.

The two-loop self-mixing evolution of eq.~\eqref{eq:partial-nll-rge}, which we call partial NLL, gives
\begin{equation}
|C_{7\gamma}^{\NP,\rm partial\text{-}NLL}(\mu_b)|
=8.3960\times10^{-4},\qquad
\frac{|C_{7\gamma}^{\rm partial\text{-}NLL}|}
{|C_{7\gamma}^{\rm LL}|}-1=-25.46\%.
\label{eq:partial-nll-validation}
\end{equation}
This shift is large because the LL value itself is small. For fixed $n_f=5$ and $\eta=\alpha_s(\mu_D)/\alpha_s(\mu_b)$, the LL value $C_{7\gamma}^{\NP,\rm eff}(\mu_b)$ is the sum of the direct term $\eta^{16/23}Q_D\,C_{8G}^{\NP}(\mu_D)$ and the mixing term $\tfrac83(\eta^{14/23}-\eta^{16/23})\,C_{8G}^{\NP}(\mu_D)$ \cite{Buchalla:1995vs}, which have opposite signs. From eqs.~\eqref{eq:matching-validation} and \eqref{eq:threshold-validation}, $C_{7\gamma}^{\NP,\rm eff}(\mu_b)/C_{8G}^{\NP}(\mu_D)=-0.071$, about five times smaller than the matching value $Q_D=-1/3$. Therefore, a moderate higher-order correction to either term is strongly amplified in their difference. It is significant to mention here that this affects a small residual WC, and not a dominant radiative amplitude.

{\emergencystretch=2em For the observables, the effect is negligible. At the profiled best-fit point, $|C_{7\gamma}^{\NP}/C_{7\gamma}^{\SM,\rm eff}|=\changed{5.3\times10^{-4}}$. Since the primed dipole enters the rate only through its square, the LL prediction for $\BR(B\to X_s\gamma)$ exceeds the SM value by \changed{$2.8\times10^{-7}$} in relative terms. The partial-NLL evolution reduces this excess by \changed{$1.3\times10^{-7}$} and changes $R_{d/s}^{\gamma}$ by \changed{$7\times10^{-9}$} in relative terms. Therefore, the fit results presented below use the LL evolution, and the missing $\mathcal O(\alpha_s)$ portal matching cannot affect any endpoint quoted in this paper.\par}

\subsection{Best fit at the one-TeV benchmark}
\label{sec:best-fit}

We now turn to the fit itself. With explicit nuisance profiling, the SM and the best complex-scalar points give
\begin{equation}
\chi^2_{\SM,\rm profile}=\changed{5.954},\qquad
\chi^2_{\rm complex,best}=\changed{0.296},\qquad
\Delta\chi^2_{\SM}=\changed{5.658}.
\label{eq:main-fit-result}
\end{equation}
The best-fit portal parameters are
\begin{equation}
\epsR=\changed{0.146},\qquad \thetaR=\changed{0.201},
\qquad \phi_d=\changed{-3.046},\qquad \phi_s=\changed{-3.019},
\label{eq:best-fit-parameters}
\end{equation}
\changed{corresponding to $\phi_s-\phi_d=0.027$, that is, to a nearly common portal phase. A relative sign between $\lambda_d$ and $\lambda_s$ does not change any of the observables considered here, see section~\ref{sec:alignment}, so that the point with $\phi_d$ shifted by $\pi$ gives the same $\chi^2$.} The associated coupling products are $|\Lambda_{sb}|=\changed{0.143}$, $|\Lambda_{db}|=\changed{0.029}$, and $|\Lambda_{ds}|=\changed{4.2\times10^{-3}}$, and the profiled predictions for the seven observables are collected in table~\ref{tab:benchmark-predictions}. The three most precisely measured observables, $\Delta M_s$, $\Delta M_d$, and $|\epsilon_K|$, are matched in both columns, because their theory uncertainties are much larger than the experimental ones and the nuisances absorb the difference.

The real scalar improves on the profiled SM by only $\Delta\chi^2=2\times10^{-4}$. This follows from the structure of the model: its boxes cancel, its dipoles enter the rate only quadratically and are small, and its $Z$ penguin vanishes while the photon penguin is negligible, see section~\ref{sec:semileptonic}. The real scalar is not blind to flavor, since the dipoles and $C_9'$ are nonzero, but it is unconstrained below the scan boundary $\epsR\leq1$ by the observables considered in this work.

\begin{table}[tbp]
\centering
\small
\begin{tabular}{@{}lccrcr@{}}
\toprule
& & \multicolumn{2}{c}{Profiled SM} &
  \multicolumn{2}{c}{Best complex scalar} \\
\cmidrule(lr){3-4}\cmidrule(lr){5-6}
Observable & Experiment & value & pull & value & pull \\
\midrule
$\Delta M_s\,[\mathrm{ps}^{-1}]$ & $17.766(6)$~\cite{HeavyFlavorAveragingGroupHFLAV:2024ctg}
  & \changed{$17.766$} & \changed{$-0.01$} & \changed{$17.766$} & \changed{$+0.00$} \\
$\Delta M_d\,[\mathrm{ps}^{-1}]$ & $0.5069(19)$~\cite{HeavyFlavorAveragingGroupHFLAV:2024ctg}
  & \changed{$0.5070$} & \changed{$+0.07$} & \changed{$0.5069$} & \changed{$-0.01$} \\
$\phi_s^{c\bar cs}\,[\mathrm{rad}]$ & $-0.052(13)$~\cite{HeavyFlavorAveragingGroupHFLAV:2024ctg}
  & \changed{$-0.0376$} & \changed{$+1.11$} & \changed{$-0.0513$} & \changed{$+0.06$} \\
{\color{changedcolor}$2\beta_{\rm eff}\,[^\circ]$} & {\color{changedcolor}$45.24(90)$}~\cite{HFLAV:Summer2025}
  & \changed{$45.92$} & \changed{$+0.76$} & \changed{$45.20$} & \changed{$-0.04$} \\
$|\epsilon_K|\,[10^{-3}]$ & $2.228(11)$~\cite{ParticleDataGroup:2024cfk}
  & \changed{$2.227$} & \changed{$-0.11$} & \changed{$2.228$} & \changed{$+0.00$} \\
$\BR(B\to X_s\gamma)\,[10^{-4}]$ & $3.49(19)$~\cite{HeavyFlavorAveragingGroupHFLAV:2024ctg}
  & $3.440$ & $-0.26$ & $3.440$ & $-0.26$ \\
$R_{d/s}^{\gamma}$ & $0.040(13)$~\cite{BaBar:2010vgu}
  & \changed{$0.0439$} & \changed{$+0.29$} & \changed{$0.0444$} & \changed{$+0.33$} \\
\midrule
$\sum(\text{pull})^2$ & & \multicolumn{2}{c}{\changed{$1.99$}}
  & \multicolumn{2}{c}{\changed{$0.18$}} \\
nuisance penalty & & \multicolumn{2}{c}{\changed{$3.97$}}
  & \multicolumn{2}{c}{\changed{$0.12$}} \\
\bottomrule
\end{tabular}
\caption{Experimental values and profiled predictions at the benchmark in eq.~\eqref{eq:mass-benchmark}. Pulls are $(O^{\rm th}-O^{\rm exp})/\sigma_{\rm exp}$; the pull and nuisance terms sum to the $\chi^2$ values in eq.~\eqref{eq:main-fit-result}.}
\label{tab:benchmark-predictions}
\end{table}

The improvement in eq.~\eqref{eq:main-fit-result} is not evidence for NP. Four portal parameters are fitted, and the phases are unidentified at the null point. In the SM pseudo-experiments, \changed{202} out of 1000 have a likelihood-ratio statistic at least as large as the observed value $q_{\SM}=\changed{5.66}$, which gives the Monte Carlo (MC) $p$-value $p_{\rm MC}=\changed{0.20}$ with a 95\% Wilson interval \changed{$[0.18,0.23]$} \cite{Wilson:1927}. The fit rather shows a correlated solution, in which the same spurion improves $\phi_s^{c\bar cs}$ and \changed{$2\beta_{\rm eff}$} without a significant kaon or radiative penalty.

Furthermore, every profiled nuisance pull at the best-fit point is smaller than \changed{$0.24$} standard deviations, see figure~\ref{fig:nuisance-pulls} in appendix~\ref{app:diagnostics}. Therefore, the improvement does not come from pushing a CKM, lattice, or SD input to the edge of its prior. \changed{In the profiled SM, in contrast, the nuisance penalty is larger than the contribution of the observables, see table~\ref{tab:benchmark-predictions}, mainly because $|V_{cb}|$ is shifted by $1.3$ standard deviations.}

\subsection{Test of the minimal right-handed flavor structure}
\label{sec:alignment}

In this subsection, we \changed{address the test of the minimal right-handed flavor structure}. The generic fit of section~\ref{sec:best-fit} uses four flavor parameters and leaves the relation between $\lambda_d$ and $\lambda_s$ free, see section~\ref{sec:portal-params}. In order to test whether the data require this freedom, we impose the minimal $U(2)^3$ relation in eq.~\eqref{eq:minimal-u2} and the CKM-aligned ansatz in eq.~\eqref{eq:ckm-alignment} as nested restrictions of the generic fit. In both cases, $\thetaR$ and $\phi_d-\phi_s$ are fixed by the CKM matrix at each nuisance point, so that only $\epsR$ and one common phase remain free. We include both signs of eqs.~\eqref{eq:minimal-u2} and \eqref{eq:ckm-alignment} in the minimization. As expected from section~\ref{sec:portal-params}, they give the same $\chi^2$, which provides a check on the minimization.

\begin{table}[tbp]
\centering
\small
\begin{tabular}{@{}lccccccc@{}}
\toprule
Hypothesis & Params & $\tan\thetaR$ & $\sum(\text{pull})^2$ & Nuisance
  & $\chi^2$ & $\epsR$ & $\epsR^{95}$ \\
\midrule
SM & $0$ & --- & \changed{$1.99$} & \changed{$3.97$} & \changed{$5.954$} & --- & --- \\
Minimal $U(2)^3$ & $2$ & $0.011$ & \changed{$0.72$} & \changed{$2.23$} & \changed{$2.948$} & \changed{$0.135$}
  & \changed{$0.211$} \\
CKM-aligned ansatz & $2$ & \changed{$0.207$} & \changed{$0.22$} & \changed{$1.56$} & \changed{$1.777$} & \changed{$0.090$}
  & $0.112$ \\
Generic $V_d$ & $4$ & fitted & \changed{$0.18$} & \changed{$0.12$} & \changed{$0.296$} & \changed{$0.146$}
  & \changed{$0.227$} \\
\bottomrule
\end{tabular}
\caption{Profiled fits of the four hypotheses at $M_D\simeq m_{\mathcal S}=1\TeV$ and $|\lambda_b|=1$; $\chi^2$ is the sum of the pull and nuisance terms.}
\label{tab:alignment-test}
\end{table}

The results are given in table~\ref{tab:alignment-test}, where all fifteen nuisances are profiled in every row and $\tan\thetaR$ is predicted in the two restricted rows. \changed{With the same number of parameters, the CKM-aligned ansatz gives a lower profiled $\chi^2$ than minimal $U(2)^3$, $1.78$ against $2.95$. Table~\ref{tab:alignment-test} shows that it describes the seven observables better, with $\sum(\text{pull})^2=0.22$ against $0.72$, and that it also requires a smaller nuisance penalty, $1.56$ against $2.23$.} In minimal $U(2)^3$, the $m_d/m_s$ suppression in eq.~\eqref{eq:minimal-u2} makes $h_d$ negligible. Therefore, the portal corrects $\phi_s^{c\bar cs}$, which has the largest SM pull in table~\ref{tab:benchmark-predictions}, through $B_s$ mixing alone, while \changed{$2\beta_{\rm eff}$} keeps a pull of \changed{$0.69$, and the CKM inputs $|V_{cb}|$, $|V_{ub}|$, and $\gamma$ are shifted by up to $0.9$ standard deviations. In the CKM-aligned ansatz, $h_d=h_s$, so that the portal also contributes to $B_d$ mixing, and the pull of $2\beta_{\rm eff}$ is reduced to $0.29$.} Relaxing the restrictions to a generic spurion gives
\begin{equation}
\Delta\chi^2_{\rm minimal}=\chi^2_{\rm minimal}-\chi^2_{\rm generic}=\changed{2.65},
\qquad
\Delta\chi^2_{\rm aligned}=\chi^2_{\rm aligned}-\chi^2_{\rm generic}=\changed{1.48},
\label{eq:alignment-lr}
\end{equation}
with asymptotic $p$-values of \changed{$0.27$ and $0.48$}, respectively, from a $\chi^2$ distribution with two degrees of freedom, $\chi^2_2$ \cite{Wilks:1938dza}, since the generic fit has two extra parameters. Therefore, neither restriction is disfavored by the data.

The comparison between the two restrictions needs two remarks. First, they are not nested in one another and have the same number of parameters. Therefore, the difference $\chi^2_{\rm minimal}-\chi^2_{\rm aligned}=\changed{1.17}$ shows which of them describes the data better under our likelihood, but it is not a significance, and a difference of about one $\chi^2$ unit is not evidence for either. \changed{Second, this difference depends on the treatment of the CKM inputs. Table~\ref{tab:ckm-robustness} repeats the fits for three alternative treatments. In the first, the statistical and theoretical errors in eq.~\eqref{eq:ckm-inputs} are added linearly, which is closer to the Rfit treatment of ref.~\cite{ValeSilva:2024jml}. In the second, we use instead the Wolfenstein parameters $A$, $\bar\rho$, and $\bar\eta$ of the tree-level fit of ref.~\cite{ValeSilva:2024jml} as independent nuisance parameters with split-normal priors. In the third, the correlations among $A$, $\bar\rho$, and $\bar\eta$ implied by eq.~\eqref{eq:ckm-inputs} are added to the second treatment. The generic fit gives the lowest $\chi^2$ in every case, as it must, but the ordering of the two restrictions is not stable. The CKM-aligned ansatz is preferred for the inputs of eq.~\eqref{eq:ckm-inputs}, and minimal $U(2)^3$ for the tree-level Wolfenstein parameters, with $\chi^2_{\rm minimal}-\chi^2_{\rm aligned}$ ranging from $-0.91$ to $1.17$. The reason is that the SM prediction of $2\beta_{\rm eff}$ lies above the measurement, see table~\ref{tab:benchmark-predictions}, and this tension can be reduced either by lowering $\bar\eta$ through the CKM nuisances or by a portal contribution to $B_d$ mixing. Minimal $U(2)^3$ can use only the first, while the CKM-aligned ansatz can also use the second. The downward uncertainty of $\bar\eta$ is $0.023$ in the tree-level fit of ref.~\cite{ValeSilva:2024jml}, but $0.017$ when $\bar\eta$ is derived from eq.~\eqref{eq:ckm-inputs}, so that a shift of $\bar\eta$ is cheaper, and minimal $U(2)^3$ is preferred, for the tree-level Wolfenstein parameters. Therefore, the present data cannot discriminate between the two restrictions. In addition, neither restriction is disfavored for any of the four treatments, with asymptotic $p$-values between $0.27$ and $0.57$.}

\begin{table}[tbp]
\centering
\small
\begin{tabular}{@{}lcccc@{}}
\toprule
CKM treatment & SM & Minimal $U(2)^3$ & CKM-aligned & Generic \\
\midrule
{\color{changedcolor}Eq.~\eqref{eq:ckm-inputs}, quadrature (baseline)} & \changed{$5.954$} & \changed{$2.948$} & \changed{$1.777$} & \changed{$0.296$} \\
{\color{changedcolor}Eq.~\eqref{eq:ckm-inputs}, linear} & \changed{$4.592$} & \changed{$2.130$} & \changed{$1.406$} & \changed{$0.292$} \\
{\color{changedcolor}Tree-level $A$, $\bar\rho$, $\bar\eta$} & \changed{$5.710$} & \changed{$1.740$} & \changed{$2.649$} & \changed{$0.493$} \\
{\color{changedcolor}Tree-level $A$, $\bar\rho$, $\bar\eta$, correlated} & \changed{$5.254$} & \changed{$1.978$} & \changed{$2.233$} & \changed{$0.479$} \\
\bottomrule
\end{tabular}
\caption{Profiled $\chi^2$ for \changed{alternative treatments of the CKM inputs}.}
\label{tab:ckm-robustness}
\end{table}

The $p$-values above are asymptotic. Unlike the SM comparison in eq.~\eqref{eq:main-fit-result}, these likelihood ratios are evaluated at $\epsR>0$, where the two additional angles are identified, so that the asymptotic law should work better here. We check this for the CKM-aligned ansatz with 400 pseudo-experiments generated at its best-fit point. Each pseudo-experiment is fitted with the restriction imposed and released, which gives $q=\chi^2_{\rm aligned}-\chi^2_{\rm generic}$. \changed{In all 400 of them, the generic fit reaches a $\chi^2$ at least as low as the restricted one, as it must.}

The distribution of $q$ is broader than $\chi^2_2$, with a $95\%$ critical value of \changed{$7.00$} instead of $5.99$, so that the asymptotic threshold covers only \changed{$92.5\%$} of the pseudo-experiments. The calibrated $p$-value is
\begin{equation}
p_{\rm MC}^{\rm align}=\changed{0.62\quad[0.57,0.66]},
\label{eq:alignment-pvalue}
\end{equation}
with a $95\%$ Wilson interval, larger than the asymptotic value \changed{$0.48$}. Therefore, the present data give no indication that the right-handed spurion deviates from CKM alignment. \changed{The same calibration with 400 pseudo-experiments generated at the minimal $U(2)^3$ best fit, all of which satisfy nesting, gives a $95\%$ critical value of $6.75$, a coverage of $93.8\%$ for the asymptotic threshold, and
\begin{equation}
p_{\rm MC}^{\rm minimal}=0.35\quad[0.30,0.39],
\label{eq:minimal-pvalue}
\end{equation}
larger than the asymptotic value $0.27$. Therefore, the present data give no indication against the minimal direction either.} In addition, the asymptotic critical value is too small by a factor of $2.3$ at the SM boundary, see section~\ref{sec:epsilon-profile}, but only by \changed{$17\%$ and $13\%$} here, where the extra parameters are identified. Therefore, the failure of the $\chi^2$ approximation follows the loss of identifiability.

\begin{figure}[tbp]
\centering
\includegraphics[width=0.70\textwidth]{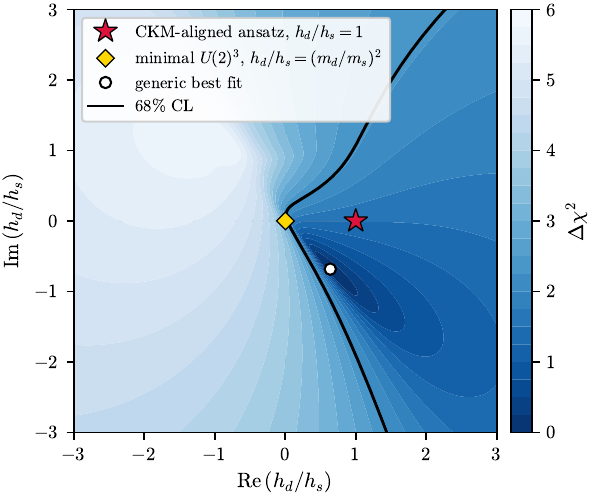}
\caption{Profiled $\Delta\chi^2$ in the complex $h_d/h_s$ plane at $M_D\simeq m_{\mathcal S}=1\TeV$; the solid line is the $68\%$ contour. The star is the CKM-aligned ansatz, the diamond is minimal $U(2)^3$, and the open circle is the generic best fit.}
\label{fig:correlation-region}
\end{figure}

The same result follows from the flavor structure. The rank-one relation means that the two $b$-sector products differ by a single complex number $w=\Lambda_{db}/\Lambda_{sb}$. Since $C_{RR}^{qb}\propto\Lambda_{qb}^2$ and $C_{LL}^{q,\SM}\propto(V_{tq}^*V_{tb})^2$, see eqs.~\eqref{eq:complex-box-coefficient} and \eqref{eq:hq-definition}, we obtain
\begin{equation}
\frac{h_d}{h_s}=w^2\left(\frac{V_{ts}^*}{V_{td}^*}\right)^2.
\label{eq:hd-over-hs}
\end{equation}
Therefore, the CKM-aligned ansatz is exactly the point $h_d/h_s=1$, and minimal $U(2)^3$ is the point $h_d/h_s=(m_d/m_s)^2\simeq2.5\times10^{-3}$, practically at the origin. In figure~\ref{fig:correlation-region}, we scan $w$ and profile $\epsR$, the common phase, and all fifteen nuisances at each point. Both signs of eqs.~\eqref{eq:minimal-u2} and \eqref{eq:ckm-alignment} map to the same point, since $h_d/h_s$ depends on $w^2$. \changed{The aligned point lies inside the $68\%$ region at $\Delta\chi^2=1.51$, and the minimal $U(2)^3$ point lies just outside it at $\Delta\chi^2=2.65$.} These values agree with eq.~\eqref{eq:alignment-lr}; the small difference for the aligned point arises because $w$ is held at its central CKM value in the scan, while in the direct fit it follows the profiled CKM matrix. The generic best fit lies near $h_d/h_s\simeq\changed{0.65-0.69\,i}$, and the $68\%$ region is open towards large $|h_d/h_s|$ in this direction.

No $95\%$ contour exists, because $\epsR\to0$ returns the SM, which is compatible with the data. \changed{Therefore, $\Delta\chi^2$ cannot exceed $\chi^2_{\SM,\rm profile}-\chi^2_{\rm complex,best}=5.66$ anywhere in the plane, which is below the two-parameter $95\%$ threshold of $5.99$.} Therefore, every value of $h_d/h_s$ is allowed at this level, and the flavor structure of the portal can be constrained only once its overall strength is known to be nonzero.

Finally, both restrictions constrain $\epsR$ more strongly than the generic case. Profiling $\epsR$ under eqs.~\eqref{eq:minimal-u2} and \eqref{eq:ckm-alignment} gives, see also table~\ref{tab:alignment-test},
\begin{equation}
\epsR^{95,\rm minimal}=\changed{0.211},\qquad
\epsR^{95,\rm aligned}=0.112,
\label{eq:epsilon-upper-restricted}
\end{equation}
against the generic endpoint of eq.~\eqref{eq:epsilon-upper-benchmark}. Therefore, the present data give $\epsR\leq0.112$ if right-handed flavor breaking is CKM-aligned, $\epsR\leq\changed{0.211}$ if it is minimal, and $\epsR\leq\changed{0.227}$ if the spurion is generic, at $95\%$ CL. The aligned ansatz is the most constraining, because its larger $\tan\thetaR\simeq0.21$ puts a larger portal coupling into the $b\to d$ sector, where $\Delta M_d$ and \changed{$2\beta_{\rm eff}$} constrain it. The two restricted endpoints use the asymptotic threshold of eq.~\eqref{eq:profile-threshold}; only the generic endpoint is coverage-calibrated, see section~\ref{sec:epsilon-profile}.

\changed{For minimal $U(2)^3$, the best fit $\epsR=0.135$ and the upper limit $\epsR^{95,\rm minimal}=0.211$ are larger than the natural size in eq.~\eqref{eq:natural-epsilon} by factors of about $180$ and $280$, respectively. A portal effect of the fitted size with the minimal direction therefore requires the product of the coefficient in eq.~\eqref{eq:minimal-vd} and $|\lambda_V|/|\lambda_b|$ to be of this order. Therefore, the restricted fit tests the direction of the light-quark couplings predicted by minimal $U(2)^3$ with a free normalization, and not the full minimal prediction, which is indistinguishable from the SM.}

\subsection{Upper limit on the light-quark couplings}
\label{sec:epsilon-profile}

\begin{figure}[tbp]
\centering
\includegraphics[width=0.80\textwidth]{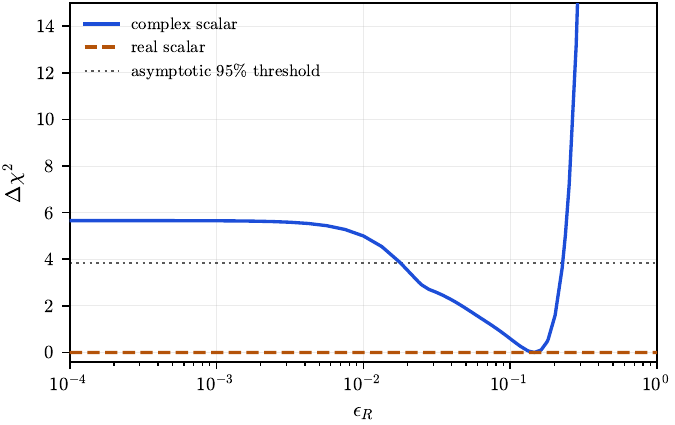}
\caption{Profile in $\epsR$ for the complex and real scalar; the dotted line is the asymptotic 95\% threshold.}
\label{fig:epsilon-profile}
\end{figure}

In this subsection, we present the profile in $\epsR$. Figure~\ref{fig:epsilon-profile} shows the one-dimensional profile, where $\thetaR$, both phases, and all fifteen nuisances are minimized at each value of $\epsR$. For the complex scalar, the asymptotic upper endpoint defined by eq.~\eqref{eq:profile-threshold} is
\begin{equation}
\epsR^{95}=\changed{0.227}
\qquad (M_D=m_{\mathcal S}=1\TeV).
\label{eq:epsilon-upper-benchmark}
\end{equation}
For small $\epsR$, the profile approaches $\Delta\chi^2_{\SM}=\changed{5.66}$ of eq.~\eqref{eq:main-fit-result}, and it crosses the threshold at $\epsR\simeq\changed{0.018}$. \changed{Between this crossing and the best fit, the profile decreases monotonically.}

Table~\ref{tab:coverage-calibration} summarizes the calibration of section~\ref{sec:scansetup}, with 1000 pseudo-experiments at each of the three truth points, and the distributions are shown in figure~\ref{fig:coverage-calibration} of appendix~\ref{app:diagnostics}. At the upper endpoint, the local 95\% critical value is $c_{0.95}=\changed{4.05}$, with a percentile-bootstrap interval \changed{$[3.55,4.45]$} \cite{Efron:1979}. Replacing the threshold in eq.~\eqref{eq:profile-threshold} by this value gives
\begin{equation}
\epsR^{95,\rm cal}=\changed{0.229},
\label{eq:epsilon-upper-calibrated}
\end{equation}
a shift of \changed{$+0.8\%$}. At this point, the threshold of $3.841$ covers \changed{$94.2\%$} of the pseudo-experiments, so the asymptotic threshold is adequate for the upper endpoint. At $\epsR=0$, however, it covers only \changed{$59.6\%$}, and the empirical critical value is \changed{$9.00$}, more than twice the asymptotic value. Therefore, the crossing of the profile at $\epsR\simeq\changed{0.018}$ cannot be given a $\chi^2_1$ significance, and it is not a lower bound on $\epsR$. \changed{All pseudo-experiments satisfy likelihood nesting, and all fits meet the convergence criterion.}

\begin{table}[tbp]
\centering
\small
\begin{tabular}{@{}lccc@{}}
\toprule
Truth point & $\epsR^{\rm true}$ & Coverage at $3.841$ & $c_{0.95}$ \\
\midrule
SM boundary & $0$ & \changed{$0.596\ [0.565,0.626]$} & \changed{$9.00\ [8.23,9.65]$} \\
Observed best fit & \changed{$0.146$} & \changed{$0.935\ [0.918,0.949]$} & \changed{$4.31\ [3.84,4.86]$} \\
Upper endpoint & \changed{$0.227$} & \changed{$0.942\ [0.926,0.955]$} & \changed{$4.05\ [3.55,4.45]$} \\
\bottomrule
\end{tabular}
\caption{Coverage calibration at three truth points, with 95\% Wilson intervals for the coverage and percentile-bootstrap intervals for $c_{0.95}$.}
\label{tab:coverage-calibration}
\end{table}

Finally, for the real scalar, the profile stays below the threshold for all $\epsR\leq1$. Its box contributions cancel, see eq.~\eqref{eq:real-box-cancellation}, so that only the radiative observables are sensitive to it, and they depend quadratically on the small dipole WCs. Therefore, the real scalar has no upper endpoint below the scan boundary.

\subsection{Mass dependence of the upper limit}
\label{sec:mass-plane}

\begin{figure}[tbp]
\centering
\includegraphics[width=0.80\textwidth]{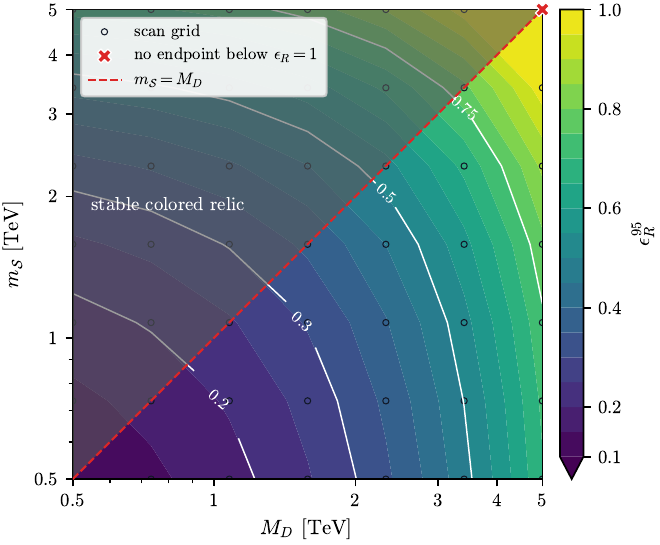}
\caption{Asymptotic 95\% upper endpoint on $\epsR$ for the complex scalar with $|\lambda_b|=1$. The shaded region $m_{\mathcal S}\geq M_D$ has a stable colored relic; crosses mark points with no endpoint below $\epsR=1$.}
\label{fig:mass-plane-limit}
\end{figure}

Figure~\ref{fig:mass-plane-limit} shows the complex-scalar upper endpoint on the mass grid of eq.~\eqref{eq:mass-plane-range}. It weakens monotonically as either mass increases, because $C_{RR}\propto G_{\rm box}(x)/M_D^2$ in eq.~\eqref{eq:complex-box-coefficient} decreases with both masses. Twenty-one of the forty-nine grid points lie in the viable region $m_{\mathcal S}<M_D$, see section~\ref{sec:nonflavor}. \changed{At all viable points an endpoint exists, and}
\begin{equation}
\changed{0.140}\leq\epsR^{95}\leq\changed{0.981},
\label{eq:mass-plane-limit-range}
\end{equation}
the strongest being at $(M_D,m_{\mathcal S})=(0.73,0.5)\TeV$ and the weakest at \changed{$(5.0,3.4)\TeV$}.

\changed{The weakest endpoint lies close to the cap $\epsR\leq1$. At the only grid point without an endpoint, $(M_D,m_{\mathcal S})=(5.0,5.0)\TeV$, the profile does not cross the threshold below $\epsR=1$, but this point lies outside the viable region.} The $\Delta F=1$ observables do not \changed{strengthen the limit at large masses}. Since the dipole enters the rate quadratically, its effect falls as $1/M_D^4$, faster than the mixing amplitudes, which are linear in $C_{RR}\propto1/M_D^2$. Furthermore, the mass-plane endpoints are asymptotic, since the pseudo-experiment calibration is performed only at the one-TeV benchmark.

This plane is flavor-only: collider, dark-matter, vacuum-stability, and perturbative-unitarity constraints are not included, see section~\ref{sec:nonflavor}.

\subsection{Dark-matter and collider constraints}
\label{sec:nonflavor}

As discussed in section~\ref{sec:model}, the $Z_2$ parity that forbids the mixing of $D$ with the SM down quarks also makes the lighter of the two new particles stable. Therefore, the benchmark is subject to cosmological and collider constraints in addition to the flavor ones. We summarize them here without a quantitative combination, which would require the dark-sector analyses of refs.~\cite{Giacchino:2015hvk,Baek:2017ykw,Biondini:2019int,Arina:2020tuw}.

Only $m_{\mathcal S}<M_D$ is viable. For $m_{\mathcal S}\geq M_D$, $D$ is the lightest $Z_2$-odd state, and a stable colored relic in the mass range of figure~\ref{fig:mass-plane-limit} is excluded by searches for anomalous heavy isotopes, which cover masses from $0.1$ to $10\TeV$ \cite{Hemmick:1989ns}. This region is shaded in figure~\ref{fig:mass-plane-limit}, and the one-TeV benchmark is understood with $m_{\mathcal S}$ slightly below $M_D$, so that $D\to q\,\mathcal S$ is allowed.

For $m_{\mathcal S}<M_D$, the scalar is stable. For a real scalar, the annihilation into light-quark pairs through $t$-channel $D$ exchange vanishes in both the $s$ and $p$ waves, and a thermal relic abundance requires either coannihilation with a near-degenerate $D$ or the gluon channels $\mathcal S\mathcal S\to gg$ and $q\bar qg$ \cite{Giacchino:2015hvk,Biondini:2019int}. For a complex scalar, the $p$ wave does not vanish \cite{Giacchino:2015hvk}. In both cases, the one-TeV benchmark lies in the coannihilation regime, so it is not obviously excluded, but the relic density constrains $|\lambda_b|$, which the flavor observables do not.

Spin-independent direct detection proceeds at tree level through $D$ exchange with valence $d$ quarks, and therefore scales as $|\lambda_d|^4$, which is small at the best-fit point. The loop-induced coupling to gluons, driven by $|\lambda_b|$, is expected to be important as well \cite{Baek:2017ykw}. However, near degeneracy the scalar-nucleon coupling is resonantly enhanced \cite{Giacchino:2015hvk}, so that direct detection can be relevant for the one-TeV benchmark. We do not evaluate these rates here. Therefore, the benchmark should be read as a flavor benchmark whose dark-sector viability is not tested in this work.

Finally, at the LHC, pair production of the colored $D$ followed by $D\to q\,\mathcal S$ gives jets plus missing energy. Since $|\lambda_{d,s}|\leq\epsR|\lambda_b|$, the decay $D\to b\,\mathcal S$ dominates, which gives the final state of bottom-squark searches. For a real scalar and a colored fermion coupled to a light quark, such searches constrain the region of small $m_{\mathcal S}$ at low $M_D$, and they weaken towards the degenerate region, where the jets are soft \cite{Arina:2020tuw}. Therefore, a combined flavor and collider analysis would remove part of the low-mass region of figure~\ref{fig:mass-plane-limit}, but not the high-mass region.

\section{Conclusions}
\label{sec:conclusions}

In this study, we have tested minimal $U(2)^3$ right-handed flavor breaking in a singlet-scalar vector-like-quark portal against a generic right-handed spurion and a CKM-aligned ansatz. Since the portal couples directly to the right-handed down quarks, and the left-handed spurion $V$ alone cannot orient a coupling to the doublet $(d_R,s_R)$, we have introduced the additional spurion $V_d$. We have computed the one-loop primed dipole WCs $C_{7\gamma}$ and $C_{8G}$ for general down-type flavors, the right-handed $\Delta F=2$ box WC $C_{RR}$, and the photon-penguin WC $C_9'$, and we have evolved them to low scales with LL running that includes the quark thresholds. Because a single mediator carries a single coupling vector, the products that control $b\to s$, $b\to d$, and $s\to d$ transitions obey the rank-one relation $\Lambda_{ds}=\Lambda_{db}\Lambda_{sb}^*/|\lambda_b|^2$, so that kaon and neutral-$B$ observables test one linked portal rather than independent couplings. Regarding the theory inputs, we have employed a profile likelihood in which the CKM, lattice, SD, and radiative inputs are fifteen nuisance parameters, minimized at every portal point.

A generic right-handed spurion predicts no relation between the $b\to d$ and $b\to s$ couplings, since its four parameters span an arbitrary complex two-vector. The testable hypothesis is minimal $U(2)^3$, in which the right-handed spurion is $\Delta Y_d^\dagger V$ and the $b\to d$ coupling is suppressed by $m_d/m_s$ relative to $b\to s$. \changed{With coefficients of order one, minimal $U(2)^3$ gives $\epsR\approx7.5\times10^{-4}$ and no visible effect, so that we have tested the direction that it predicts, with the size of the light-quark couplings left free. The best fit in this direction requires an enhancement of about $180$ over this natural size.} The current study shows that minimal $U(2)^3$, as a nested restriction with two parameters fewer than the generic fit, increases $\chi^2$ by only \changed{$2.65$}, with a \changed{calibrated} $p$-value of \changed{$0.35$}. \changed{Therefore, the minimal direction is not disfavored. However, this does not distinguish it from other directions, because the SM itself is compatible with the data, so that no direction of the light-quark couplings can be excluded at present.} The CKM-aligned ansatz $h_d=h_s$ is not disfavored either, with a calibrated $p$-value of \changed{$0.62$}. \changed{With the same number of parameters, the CKM-aligned ansatz gives a lower profiled $\chi^2$ than minimal $U(2)^3$, $1.78$ against $2.95$. However, this ordering is reversed when the tree-level Wolfenstein parameters $A$, $\bar\rho$, and $\bar\eta$ are used instead of $|V_{cb}|$, $|V_{ub}|$, and $\gamma$, because the tension in $\sin2\beta$ can be reduced either by a shift of $\bar\eta$ or by a portal contribution to $B_d$ mixing. Since the two restrictions are not nested in one another, this difference is not a significance in any case. Therefore, the present data cannot discriminate between minimal $U(2)^3$ and the CKM-aligned ansatz.} In addition, the restrictions strengthen the upper limit on the light-quark couplings at the 1 TeV benchmark, from $\epsR^{95}=\changed{0.227}$ for a generic spurion to \changed{$0.211$} for minimal $U(2)^3$ and $0.112$ for CKM alignment.

We have found that not only the size but also the structure of the constraints depends on the nature of the scalar. A complex scalar generates right-handed $\Delta F=2$ operators proportional to $(\Lambda_{ij})^2$, so that the portal phases are doubled in mixing. For a single real scalar, in contrast, the direct and crossed boxes cancel for all external flavors and all masses. This cancellation is lifted if the complex scalar is split into two real states with different masses, see appendix~\ref{app:boxcancel}. In both cases, the dipoles do not cancel, and since the photon and the gluon are emitted only from $D$, they keep the relation $C_{7\gamma}^{\NP}/C_{8G}^{\NP}=Q_D=-1/3$ at the matching scale.

At $M_D\simeq m_{\mathcal S}=1\TeV$ and $|\lambda_b|=1$, with all fifteen nuisances profiled, the profiled $\chi^2$ is \changed{$5.954$} in the SM and \changed{$0.296$} at the best complex-scalar point, which has $\epsR=\changed{0.146}$ and a nearly common portal phase. The improvement comes from $\phi_s^{c\bar cs}$ and \changed{$2\beta_{\rm eff}$}, without a significant kaon or radiative penalty. However, this improvement of $\Delta\chi^2=\changed{5.66}$ is not evidence for NP, since the pseudo-experiments give $p_{\rm MC}=\changed{0.20}$. For the radiative ratio, we have used the BaBar measurement in the window $0.5<m_X<2.0\,\mathrm{GeV}$, which avoids comparing a partial $b\to d\gamma$ rate with a fully inclusive prediction.

For the complex scalar, the asymptotic $\chi^2$ approximation fails at the SM boundary, but the pseudo-experiments show that it is adequate at the upper limit, where the coverage-calibrated limit agrees with the asymptotic value $\epsR^{95}=\changed{0.227}$ to \changed{$0.8\%$}. For $0.5\TeV\leq M_D,m_{\mathcal S}\leq5\TeV$ and in the viable region $m_{\mathcal S}<M_D$, the asymptotic upper limit weakens as either mass increases, from \changed{$0.140$ to $0.981$, and it exists at every viable grid point}. The real scalar, in contrast, has no upper limit below $\epsR=1$, since its box contributions cancel and its dipoles enter the radiative rates only quadratically. This is the main phenomenological difference between the two scalars. These limits are set by flavor data alone, and the dark-matter and collider constraints discussed in section~\ref{sec:nonflavor} are not included.

Regarding the precision of the radiative predictions, the portal contribution is evolved at LL order and applied as a rescaling of the SM rate. The partial-NLL evolution of section~\ref{sec:validation} changes the portal dipole by about $25\%$, but the radiative observables at the best-fit point by less than $2\times10^{-7}$ in relative terms. It is not a complete NLL prediction without the $\mathcal O(\alpha_s)$ portal matching and the NLO matrix elements. However, since the portal dipole enters the rate quadratically, its total effect on $\BR(B\to X_s\gamma)$ at the best-fit point is only $3\times10^{-7}$ in relative terms, so that even a correction of order one to the dipole would not change the fit. Therefore, this limitation does not affect the upper limits quoted above.

Finally, the observables that drive this analysis will improve in the coming years. The largest SM pull in our fit is that of $\phi_s^{c\bar cs}$, which will be measured more precisely at LHCb \changed{and CMS} and, in the longer term, at the HL-LHC \cite{LHCb:2018roe,Cerri:2018ypt}, while $\sin2\beta$ will be measured more precisely at both LHCb and Belle~II \cite{LHCb:2018roe,Belle-II:2018jsg}. For $\Delta M_s$, $\Delta M_d$, and $\epsilon_K$, the measurements are already much more precise than the predictions, so that their impact will grow as the theory inputs improve, in particular $f_{B_q}\sqrt{\widehat B_{B_q}}$ and the tree-level CKM parameters. Therefore, precise measurements of these observables will not only sharpen the bounds on right-handed flavor breaking in vector-like-quark portals, but also provide a window of opportunity to test whether its flavor structure is the minimal one. Since this structure can be constrained only once the portal strength is known to be nonzero, such a test requires a significant deviation from the SM in $B$-meson mixing.

\acknowledgments

The author thanks Herzallah Alharazin (Ruhr-Universit\"at Bochum) for a careful reading of the manuscript and for many valuable comments, which improved the statistical analysis, in particular the treatment of the CKM inputs as nuisance parameters, and for discussions on the natural size of the light-quark couplings and on neutral-meson mixing. He also pointed out the second solution of the $B_d$ mixing phase with $\cos2\beta_{\rm eff}<0$. \changed{The author used Claude to assist with code development and language editing.}

\section*{Data availability}

The analysis code and the numerical results from which all tables and figures of this article are obtained are provided as ancillary files with the arXiv version of this article.

\appendix
% In the appendices only the main heading is numbered; the subheadings are
% unnumbered and do not appear in the table of contents.
\addtocontents{toc}{\protect\setcounter{tocdepth}{1}}
\setcounter{secnumdepth}{1}
\section{One-loop dipole matching}
\label{app:dipole}

In this appendix, we derive the dipole WCs of eq.~\eqref{eq:dipole-matching} and the loop function $F(x)$ of eq.~\eqref{eq:dipole-loop}.

\subsubsection{Feynman rules and amplitude}

In the down-quark mass basis, the portal interaction is
\begin{equation}
\cL\supset-\lambda_i\,\Phi\,\bar DP_Rd_i-\lambda_i^*\,\Phi^\dagger\,\bar d_iP_LD,
\label{eq:app-portal}
\end{equation}
with $\Phi,\Phi^\dagger\to S_R$ for a real scalar. The vertices are $-i\lambda_iP_R$ (incoming $d_i$) and $-i\lambda_i^*P_L$ (outgoing $d_i$), and the propagators are $i(\slashed k+M_D)/(k^2-M_D^2)$ and $i/(k^2-m_{\mathcal S}^2)$. With the covariant derivative $D_\mu=\partial_\mu+ieQA_\mu+ig_sT^aG^a_\mu$, the photon and gluon couple to $D$ through $-ieQ_D\gamma^\mu$ and $-ig_sT^a\gamma^\mu$. Since $\mathcal S$ is neutral and a color singlet, these are the only gauge vertices in the loop.

For $b(p)\to q(p')\gamma(q)$, with $q=p-p'$ and the scalar carrying the loop momentum $k$, the diagram of figure~\ref{fig:dipole-diagram} gives
\begin{equation}
\begin{split}
i\mathcal M^\mu=\int\!\frac{d^dk}{(2\pi)^d}\,
&\bar u(p')(-i\lambda_q^*P_L)\frac{i(\slashed k+\slashed p'+M_D)}{(k+p')^2-M_D^2}
(-ieQ_D\gamma^\mu)\\
&\times\frac{i(\slashed k+\slashed p+M_D)}{(k+p)^2-M_D^2}
(-i\lambda_bP_R)u(p)\,\frac{i}{k^2-m_{\mathcal S}^2}.
\end{split}
\label{eq:app-dipole-amp}
\end{equation}
The factors of $i$ give $+1$, and $P_L(\slashed a+M_D)\gamma^\mu(\slashed b+M_D)P_R =\slashed a\gamma^\mu\slashed b P_R+M_D^2\gamma^\mu P_R$. The second term has no $\sigma^{\mu\nu}$ structure and does not contribute to the dipole. Hence
\begin{equation}
i\mathcal M^\mu=eQ_D\Lambda_{qb}\int\!\frac{d^dk}{(2\pi)^d}\,
\frac{\bar u(p')(\slashed k+\slashed p')\gamma^\mu(\slashed k+\slashed p)P_Ru(p)}
{(k^2-m_{\mathcal S}^2)\left[(k+p')^2-M_D^2\right]\left[(k+p)^2-M_D^2\right]}
+\cdots.
\label{eq:app-dipole-amp2}
\end{equation}

\subsubsection{Dipole projection}

With Feynman parameters $x_1$, $x_2$, and $x_3$ for the three denominators and the shift $k=\ell-x_2p'-x_3p$, the denominator becomes $(\ell^2-\Delta)^3$ with
\begin{equation}
\Delta=x_1m_{\mathcal S}^2+(1-x_1)M_D^2,
\label{eq:app-dipole-delta}
\end{equation}
where external masses are neglected against $M_D$. The terms linear in $\ell$ vanish after integration, and the terms with $\slashed\ell\gamma^\mu\slashed\ell$ are again proportional to $\gamma^\mu$. The remaining numerator is $\bar u(p')\slashed a\gamma^\mu\slashed bP_Ru(p)$ with $a=(1-x_2)p'-x_3p$ and $b=-x_2p'+(1-x_3)p$. Using $\bar u(p')\slashed p'=0$ ($m_q=0$), $\slashed pP_Ru(p)=m_bP_Lu(p)$, $q^2=0$, and $\slashed q\gamma^\mu=q^\mu+i\sigma^{\mu\nu}q_\nu$ with $\sigma^{\mu\nu}=\tfrac{i}{2}[\gamma^\mu,\gamma^\nu]$, we find
\begin{equation}
\bar u(p')\slashed a\gamma^\mu\slashed bP_Ru(p)
=-x_1x_3\,m_b\,\bar u(p')\,i\sigma^{\mu\nu}q_\nu P_Lu(p)+(\text{terms}\propto q^\mu),
\label{eq:app-dipole-dirac}
\end{equation}
where the terms proportional to $q^\mu$ vanish for a physical photon. The chirality flip occurs on the external $b$ line, through $m_b$.

\subsubsection{Loop integral}

The dipole part of the integral is finite, so that we can set $d=4$. With $\int d^4\ell/(2\pi)^4\,(\ell^2-\Delta)^{-3}=-i/(32\pi^2\Delta)$ and the factor $2$ of the Feynman parameterization,
\begin{equation}
i\mathcal M^\mu_{\rm dip}=i\,\frac{eQ_D\Lambda_{qb}\,m_b}{16\pi^2}\,
\mathcal I\;\bar u(p')\,i\sigma^{\mu\nu}q_\nu P_Lu(p),
\qquad
\mathcal I=\int_0^1\!dx_1\int_0^{1-x_1}\!dx_3\,\frac{x_1x_3}{\Delta}.
\label{eq:app-dipole-result}
\end{equation}
The $x_3$ integration gives $(1-x_1)^2/2$. With $z=1-x_1$ and $x=m_{\mathcal S}^2/M_D^2$,
\begin{equation}
\mathcal I=\frac{1}{2M_D^2}\int_0^1\!dz\,\frac{z^2(1-z)}{x+(1-x)z}.
\label{eq:app-dipole-I}
\end{equation}
The substitution $t=x+(1-x)z$ gives
\begin{equation}
\int_0^1\!dz\,\frac{z^2(1-z)}{x+(1-x)z}
=\frac{1}{(1-x)^4}\int_x^1\!dt\,\frac{(t-x)^2(1-t)}{t}
=\frac{\frac16-x+\frac{x^2}{2}+\frac{x^3}{3}-x^2\ln x}{(1-x)^4},
\label{eq:app-dipole-t}
\end{equation}
so that $\mathcal I=F(x)/M_D^2$, with $F(x)$ of eq.~\eqref{eq:dipole-loop}.

\subsubsection{Matching}

For an outgoing photon, $F_{\mu\nu}\to i(q_\mu\varepsilon^*_\nu-q_\nu\varepsilon^*_\mu)$, so that
\begin{equation}
\bar q\sigma^{\mu\nu}P_Lb\,F_{\mu\nu}\to
-2\,\bar u(p')\,i\sigma^{\mu\nu}q_\nu P_Lu(p)\,\varepsilon^*_\mu .
\end{equation}
With $\cL_{\rm eff}=-\cH_{\rm eff}$, the effective Hamiltonian of eq.~\eqref{eq:heff-dipole} then gives
\begin{equation}
i\mathcal M^\mu_{\rm eff}=-2i\,\frac{4\GF}{\sqrt2}V_{tb}V_{tq}^*\,
\Cseven^{qb}\,\frac{e\,m_b}{16\pi^2}\,
\bar u(p')\,i\sigma^{\mu\nu}q_\nu P_Lu(p).
\label{eq:app-dipole-eft}
\end{equation}
Equating eqs.~\eqref{eq:app-dipole-result} and \eqref{eq:app-dipole-eft},
\begin{equation}
\Cseven^{qb,\NP}(\mu_D)=
-\frac{\Lambda_{qb}}{4\sqrt2\GF V_{tb}V_{tq}^*M_D^2}\,Q_DF(x).
\label{eq:app-c7}
\end{equation}
For the gluon, $eQ_D\to g_sT^a$ in eq.~\eqref{eq:app-dipole-amp} and $e\to g_s$, $\sigma^{\mu\nu}\to\sigma^{\mu\nu}T^a$ in the operator, so that
\begin{equation}
\Ceight^{qb,\NP}(\mu_D)=
-\frac{\Lambda_{qb}}{4\sqrt2\GF V_{tb}V_{tq}^*M_D^2}\,F(x),
\qquad
\frac{\Cseven^{qb,\NP}(\mu_D)}{\Ceight^{qb,\NP}(\mu_D)}=Q_D.
\label{eq:app-c8}
\end{equation}
These are eqs.~\eqref{eq:dipole-matching} and \eqref{eq:dipole-ratio}. The same diagram, with the same result, applies to a real scalar.

\subsubsection{Photon penguin}

For an off-shell photon, the $\gamma^\mu$ part of the full amplitude in eq.~\eqref{eq:app-dipole-amp}, including the $M_D^2\gamma^\mu P_R$ term, is needed at order $q^2$, with massless external quarks. With the same Feynman parameters, the denominator is now $\Delta=\Delta_0-x_2x_3q^2$, where $\Delta_0$ is the $\Delta$ of eq.~\eqref{eq:app-dipole-delta}. The numerator gives $\left[\frac{2-d}{d}\ell^2+M_D^2+x_2x_3q^2\right]\bar u(p')\gamma^\mu P_Ru(p)$, up to the dipole and terms proportional to $q^\mu$. The $\ell^2$ term depends on $q^2$ through $\ln\Delta$, and at order $q^2$
\begin{equation}
i\Gamma^\mu=-i\,\frac{eQ_D\Lambda_{qb}}{16\pi^2}\,q^2\,\mathcal J\;
\bar u(p')\gamma^\mu P_Ru(p),\qquad
\mathcal J=\int_0^1\!dx_1\int_0^{1-x_1}\!dx_3\,x_2x_3
\left[\frac{2}{\Delta_0}+\frac{M_D^2}{\Delta_0^2}\right],
\label{eq:app-penguin-vertex}
\end{equation}
with $x_2=1-x_1-x_3$. The $q^2$-independent terms cancel against the self-energy diagrams, as required by gauge invariance. The $x_3$ integration gives $(1-x_1)^3/6$, and with $z=1-x_1$,
\begin{equation}
\mathcal J=\frac{1}{M_D^2}\int_0^1\!dz\,\frac{z^3}{6}
\left[\frac{2}{x+(1-x)z}+\frac{1}{\left(x+(1-x)z\right)^2}\right]
=\frac{G_9(x)}{M_D^2},
\label{eq:app-penguin-integral}
\end{equation}
with $G_9$ of eq.~\eqref{eq:penguin-loop}. This vertex agrees with the charge-radius form factor of ref.~\cite{Hisano:1995cp} for a loop with a charged fermion and a neutral scalar. The photon propagator $-ig_{\mu\nu}/q^2$ and the lepton vertex $+ie\gamma^\nu$ then give the four-fermion amplitude
\begin{equation}
i\mathcal M=-i\,\frac{e^2Q_D\Lambda_{qb}}{16\pi^2}\,\mathcal J\,
\left[\bar u(p')\gamma^\mu P_Ru(p)\right]\left[\bar u_\ell\gamma_\mu v_\ell\right],
\label{eq:app-penguin-amp}
\end{equation}
and matching to $\cH_{\rm eff}\supset-\frac{4\GF}{\sqrt2}V_{tb}V_{tq}^* C_9^{\prime\,qb}\cO_9^{\prime\,qb}$ gives eq.~\eqref{eq:c9-prime}.

\section{Loop functions and stable limits}
\label{app:loops}

\begin{figure}[tbp]
\centering
\includegraphics[width=\textwidth]{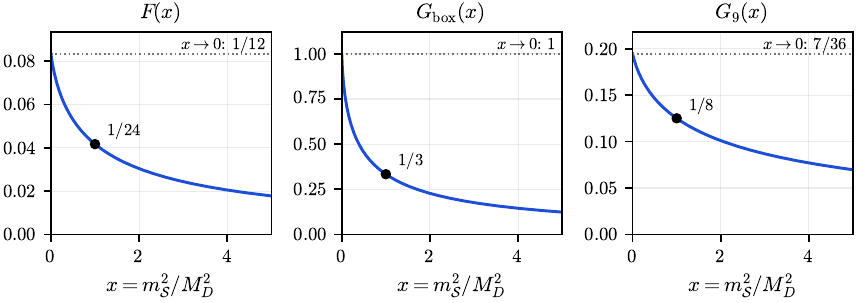}
\caption{Loop functions $F(x)$, $G_{\rm box}(x)$, and $G_9(x)$. The dots mark the values at $x=1$, and the dotted lines the limits $x\to0$.}
\label{fig:loop-functions}
\end{figure}

In this appendix, we collect the three loop functions of this work and their expansions around the degenerate point $x=m_{\mathcal S}^2/M_D^2=1$. The dipole function $F(x)$ and the box function $G_{\rm box}(x)$ are also derived and discussed in ref.~\cite{Salam:2026vys}, where further details can be found. The three functions are shown in figure~\ref{fig:loop-functions}. All of them decrease monotonically with $x$, so that a heavier scalar suppresses the portal contributions.

The dipole function can be represented as
\begin{equation}
F(x)=\frac12\int_0^1dz\,
\frac{z^2(1-z)}{z+x(1-z)}.
\label{eq:dipole-integral}
\end{equation}
Its closed form is eq.~\eqref{eq:dipole-loop}. Near the degenerate point, writing $u=x-1$, the numerically stable expansion is
\begin{equation}
F(1+u)=\sum_{n\geq0}\frac{(-1)^nu^n}{(n+2)(n+3)(n+4)}
=\frac1{24}-\frac{u}{60}+\frac{u^2}{120}
-\frac{u^3}{210}+\frac{u^4}{336}-\cdots.
\label{eq:dipole-series}
\end{equation}
In particular, $F(1)=1/24$ and $F(0)=1/12$.

For the box function,
\begin{equation}
G_{\rm box}(x)=2\int_0^1dz\,
\frac{z(1-z)}{z+x(1-z)},
\label{eq:box-integral}
\end{equation}
and
\begin{equation}
G_{\rm box}(1+u)=\sum_{n\geq0}\frac{2(-1)^nu^n}{(n+2)(n+3)}
=\frac13-\frac{u}{6}+\frac{u^2}{10}
-\frac{u^3}{15}+\frac{u^4}{21}-\cdots,
\label{eq:box-series}
\end{equation}
so that $G_{\rm box}(1)=1/3$ and $G_{\rm box}(0)=1$. The photon-penguin function of eq.~\eqref{eq:c9-prime}, with the same argument $x=m_{\mathcal S}^2/M_D^2$, is
\begin{subequations}
\label{eq:penguin-loop}
\begin{align}
G_9(x)&=\frac{-16x^3+45x^2-36x+7+6(2x-3)x^2\ln x}{36(x-1)^4},\\
G_9(1+u)&=\sum_{n\geq0}\frac{(-1)^nu^n}{(n+1)(n+2)(n+4)},
\end{align}
\end{subequations}
so that $G_9(1)=1/8$ and $G_9(0)=7/36$.

Near degeneracy, each numerator vanishes as a high power of $u=x-1$, while its individual terms remain of order unity. Therefore, the closed forms lose precision in double-precision arithmetic. For $F(x)$, the relative error is about $10^{-7}$ at $|u|=10^{-2}$ and $10^{-3}$ at $|u|=10^{-3}$. The numerical implementation therefore switches to the series for $|u|<5\times10^{-2}$. At $|u|=5\times10^{-2}$, the closed forms are accurate to better than $10^{-9}$, and the series to machine precision.
\section{Phase and Hamiltonian conventions}
\label{app:conventions}

In this appendix, we give the phase conventions and derive the sign of the portal contribution to $\epsilon_K$. The radiative Hamiltonian in eq.~\eqref{eq:heff-dipole} contains the full complex factor $V_{tb}V_{tq}^*$. Therefore, the portal phases are defined relative to the standard phase convention of the CKM matrix \cite{ParticleDataGroup:2024cfk}.

For kaons we use the operator
\begin{equation}
\cO_{RR}^{sd}=(\bar s\gamma_\mu P_Rd)(\bar s\gamma^\mu P_Rd),
\label{eq:kaon-operator-convention}
\end{equation}
whose WC is proportional to $(\lambda_s^*\lambda_d)^2=(\Lambda_{ds}^*)^2$. This operator annihilates the $d$ and $\bar s$ quarks of $K^0$ and creates the $s$ and $\bar d$ quarks of $\bar K^0$, so that $\langle\bar K^0|\cO_{RR}^{sd}|K^0\rangle\neq0$, and eq.~\eqref{eq:np-kaon-m12} defines $M_{12}^{K,\NP}$ without complex conjugation. In ref.~\cite{Buras:2010pza},
\begin{equation}
\epsilon_K=\frac{\kappa_\epsilon e^{i\phi_\epsilon}}{\sqrt2\,\Delta M_K}
\operatorname{Im}M_{12},\qquad
\operatorname{Im}M_{12}=\frac{1}{2M_K}\operatorname{Im}
\langle\bar K^0|\cH_{\rm eff}^{\Delta F=2}|K^0\rangle^*,
\label{eq:app-epsilon-k}
\end{equation}
so that $\operatorname{Im}M_{12}=-\operatorname{Im}M_{12}^{K,\NP}$ for the portal contribution. Since only $|\epsilon_K|$ enters the fit, the phase factor $e^{i\phi_\epsilon}$ drops out, and the deviation of $\phi_\epsilon$ from $45^\circ$ is included in $\kappa_\epsilon$ \cite{Buras:2010pza}. This gives eq.~\eqref{eq:np-epsilon-k}. The SM amplitude is written in the same convention, so that the relative sign of the SM and portal contributions to $\epsilon_K$ is fixed.

Finally, the conventional charm-top and the $u$-$t$ expressions for the SM kaon amplitude are algebraically equivalent, since they are related by the unitarity of the CKM matrix. Their predictions differ only through the different SD QCD factors, $\eta_{cc}$, $\eta_{ct}$, and $\eta_{tt}$ in the charm-top form, and $\eta_{ut}$ and $\eta_{tt}$ in the $u$-$t$ form \cite{Brod:2019rzc}.
\section{The \texorpdfstring{$\Delta F=2$}{Delta F=2} box diagrams}
\label{app:boxcancel}

In this appendix, we derive the complex-scalar WC of eq.~\eqref{eq:complex-box-coefficient} and the real-scalar cancellation of eq.~\eqref{eq:real-box-cancellation}, and we show how the cancellation is lifted for two real scalars with different masses.

\subsubsection{Feynman rules and amplitudes}

We use the interaction, vertices, and propagators of eq.~\eqref{eq:app-portal} in appendix~\ref{app:dipole}. We consider $b(p_1)\,b(p_2)\to q(p_3)\,q(p_4)$ at zero external momenta, with the $D$ lines running from $x_1$ to $x_3$ and from $x_2$ to $x_4$, and loop momentum $k$ on the first $D$ line.

In the direct diagram, the scalars join $x_1$ with $x_4$ and $x_2$ with $x_3$, and both $D$ lines carry $k$:
\begin{equation}
\begin{split}
i\mathcal M_{\rm dir}=\int\!\frac{d^4k}{(2\pi)^4}\,
&\left[\bar u_3(-i\lambda_q^*P_L)\frac{i(\slashed k+M_D)}{k^2-M_D^2}
(-i\lambda_bP_R)u_1\right]\\
&\times\left[\bar u_4(-i\lambda_q^*P_L)\frac{i(\slashed k+M_D)}{k^2-M_D^2}
(-i\lambda_bP_R)u_2\right]
\left(\frac{i}{k^2-m_{\mathcal S}^2}\right)^{\!2}.
\end{split}
\label{eq:app-amp-direct}
\end{equation}
Using $P_L(\slashed k+M_D)P_R=\slashed kP_R$ and $\Lambda_{qb}=\lambda_q^*\lambda_b$,
\begin{equation}
i\mathcal M_{\rm dir}=\Lambda_{qb}^2\,
[\bar u_3\gamma^\mu P_Ru_1][\bar u_4\gamma^\nu P_Ru_2]\,I_{\mu\nu},\qquad
I_{\mu\nu}=\int\!\frac{d^4k}{(2\pi)^4}\,
\frac{k_\mu k_\nu}{(k^2-M_D^2)^2(k^2-m_{\mathcal S}^2)^2}.
\label{eq:app-amp-direct-reduced}
\end{equation}
For a real scalar, the crossed diagram joins $x_1$ with $x_2$ and $x_3$ with $x_4$. The second $D$ line then carries $-k$, with identical denominators and no relative sign from Fermi statistics, so that
\begin{equation}
i\mathcal M_{\rm cross}=-\,i\mathcal M_{\rm dir}.
\label{eq:app-amp-crossed}
\end{equation}

\subsubsection{Loop integral and WC}

With $I_{\mu\nu}=\tfrac14 g_{\mu\nu}I_\rho^{\ \rho}$ and a Wick rotation, $t=k_E^2$,
\begin{equation}
I_{\mu\nu}=-\frac{i\,g_{\mu\nu}}{64\pi^2}\,J,\qquad
J=\int_0^\infty\!dt\,\frac{t^2}{(t+M_D^2)^2(t+m_{\mathcal S}^2)^2}.
\label{eq:app-loop-integral}
\end{equation}
With $a=M_D^2$, $b=m_{\mathcal S}^2$, and $x=b/a$,
\begin{equation}
J=\frac{a+b}{(a-b)^2}-\frac{2ab\ln(a/b)}{(a-b)^3}
=\frac{1}{M_D^2}\,\frac{1-x^2+2x\ln x}{(1-x)^3}
=\frac{G_{\rm box}(x)}{M_D^2},
\label{eq:app-J}
\end{equation}
with $G_{\rm box}$ of eq.~\eqref{eq:box-loop}. Hence
\begin{equation}
i\mathcal M_{\rm dir}=-\,i\,\frac{\Lambda_{qb}^2}{64\pi^2M_D^2}\,
G_{\rm box}(x)\,
[\bar u_3\gamma^\mu P_Ru_1][\bar u_4\gamma_\mu P_Ru_2].
\label{eq:app-amp-result}
\end{equation}
For a complex scalar, this is the only contribution. Matching onto $\cH_{\rm eff}=\CRR^{qb}\cO_{RR}^{qb}+{\rm h.c.}$, whose two identical currents give a factor of $2$ in the matrix element, yields
\begin{equation}
\CRR^{qb}(\mu_D)=\frac{\Lambda_{qb}^2}{128\pi^2M_D^2}\,G_{\rm box}(x),
\label{eq:app-crr}
\end{equation}
which is eq.~\eqref{eq:complex-box-coefficient}.

\subsubsection{Cancellation for a single real scalar}

From eq.~\eqref{eq:app-amp-crossed},
\begin{equation}
i\mathcal M_{\rm dir}+i\mathcal M_{\rm cross}=0,\qquad
\CRR^{qb,\,\rm real}=0,
\label{eq:app-real-cancel}
\end{equation}
for all $M_D$, $m_{\mathcal S}$, and external flavors. This agrees with ref.~\cite{Arnan:2019uhr}, where the crossed box of a real scalar is the direct one with the opposite sign.

\subsubsection{Two non-degenerate real scalars}
\label{app:split}

After electroweak symmetry breaking, the terms of eq.~\eqref{eq:u1-violating-terms} combine into $\mu^2\Phi^2+{\rm h.c.}$, with $\mu^2$ including $\lambda_2'v^2/2$. With $\Phi=(S_1+iS_2)/\sqrt2$ and the phase of $\Phi$ chosen such that $\mu^2=-|\mu^2|$,
\begin{equation}
m_{\mathcal S}^2|\Phi|^2+\left(\mu^2\Phi^2+{\rm h.c.}\right)
=\tfrac12\left(m_{\mathcal S}^2-2|\mu^2|\right)S_1^2
+\tfrac12\left(m_{\mathcal S}^2+2|\mu^2|\right)S_2^2,
\label{eq:app-split-masses}
\end{equation}
so that $m_{1,2}^2=m_{\mathcal S}^2\mp2|\mu^2|$. With $\Delta_{1,2}=i/(k^2-m_{1,2}^2)$,
\begin{equation}
\langle\Phi\Phi^\dagger\rangle=\tfrac12(\Delta_1+\Delta_2),\qquad
\langle\Phi\Phi\rangle=\langle\Phi^\dagger\Phi^\dagger\rangle
=\tfrac12(\Delta_1-\Delta_2).
\label{eq:app-split-propagators}
\end{equation}
The factor $\left(i/(k^2-m_{\mathcal S}^2)\right)^2$ in eq.~\eqref{eq:app-amp-direct} is then replaced by the sum of the direct and crossed contributions, the latter with the opposite numerator sign,
\begin{equation}
\tfrac14(\Delta_1+\Delta_2)^2-\tfrac14(\Delta_1-\Delta_2)^2=\Delta_1\Delta_2
=-\frac{1}{(k^2-m_1^2)(k^2-m_2^2)},
\label{eq:propagator-product}
\end{equation}
so that the loop integral of eq.~\eqref{eq:app-amp-direct-reduced} becomes
\begin{equation}
I^{(12)}_{\mu\nu}=\int\!\frac{d^4k}{(2\pi)^4}\,
\frac{k_\mu k_\nu}{(k^2-M_D^2)^2(k^2-m_1^2)(k^2-m_2^2)} .
\label{eq:app-split-integral}
\end{equation}
Following eqs.~\eqref{eq:app-loop-integral}-\eqref{eq:app-crr},
\begin{equation}
\CRR^{qb}(\mu_D)=\frac{\Lambda_{qb}^2}{128\pi^2}\,J_{12},\qquad
J_{12}=\int_0^\infty\!dt\,\frac{t^2}{(t+M_D^2)^2(t+m_1^2)(t+m_2^2)}.
\label{eq:app-split-crr}
\end{equation}
{\emergencystretch=2em For $m_1=m_2=m_{\mathcal S}$, eq.~\eqref{eq:app-crr} is recovered. For $m_2\to\infty$, in contrast, $J_{12}\simeq\ln(m_2^2/M_D^2)/m_2^2\to0$, which is the single-real-scalar limit of eq.~\eqref{eq:app-real-cancel}.\par}

\section{Fit diagnostics}
\label{app:diagnostics}

\begin{figure}[htbp]
\centering
\includegraphics[width=0.62\textwidth]{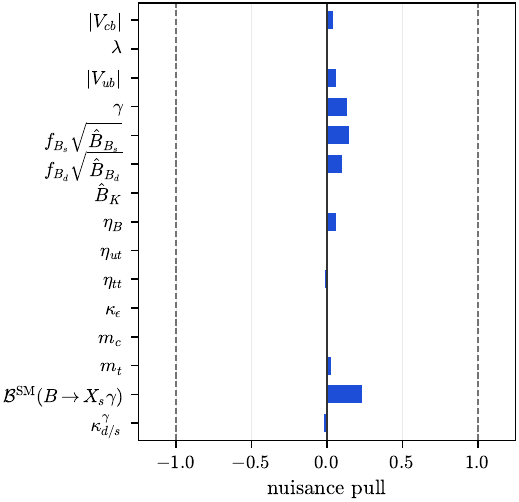}
\caption{Profiled theory-nuisance pulls at the best complex-scalar point for $M_D=m_{\mathcal S}=1\TeV$ and $|\lambda_b|=1$. The dashed lines mark $\pm1$ standard deviation.}
\label{fig:nuisance-pulls}
\end{figure}

\begin{figure}[htbp]
\centering
\includegraphics[width=0.80\textwidth]{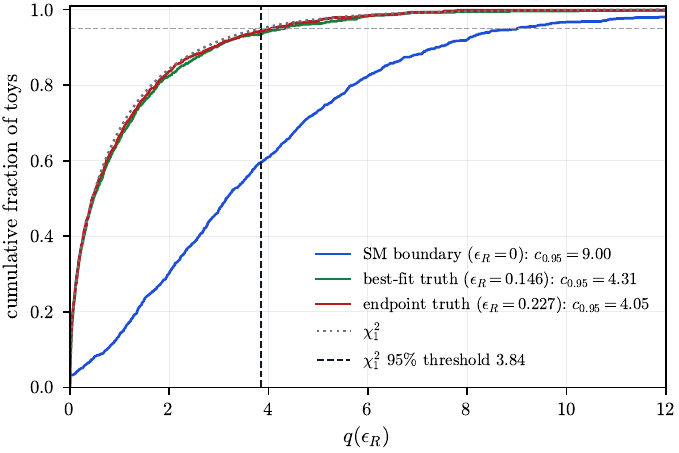}
\caption{Empirical cumulative distributions of the profile-likelihood statistic $q(\epsR)$ at the three calibration points. The dotted curve is the $\chi^2_1$ distribution; the vertical dashed line marks $3.841$ and the horizontal dashed line marks 95\%.}
\label{fig:coverage-calibration}
\end{figure}

In this appendix, we show two diagnostics of the fit. Figure~\ref{fig:nuisance-pulls} shows the profiled nuisance pulls at the best complex-scalar point. All of them are smaller than \changed{$0.24$} standard deviations, so that the fit does not rely on shifting any input far from its central value. Figure~\ref{fig:coverage-calibration} shows the distributions of the profile-likelihood statistic $q(\epsR)$ at the three calibration points of table~\ref{tab:coverage-calibration}. At the SM boundary, the distribution is much broader than $\chi^2_1$, while at the best fit and at the upper limit it follows $\chi^2_1$ closely.

\FloatBarrier

%\bibliographystyle{JHEP}
%\bibliography{references}

\providecommand{\href}[2]{#2}\begingroup\raggedright\endgroup

\end{document}